\documentclass[11pt]{amsart}

\usepackage{graphicx}
\usepackage{psfrag}
\usepackage{amsmath,bbm} 
\usepackage{amsfonts} 
\usepackage{amssymb}
\usepackage{stmaryrd}
\usepackage{array}

\def\N{{\mathbb{N}}}
\def\R{{\mathbb{R}}}
\def\Z{{\mathbb{Z}}}

\def\Div{\textup{div\,}}

\newtheorem{thm}{Theorem}
\newtheorem{lem}{Lemma}
\newtheorem{proposition}{Proposition}
\newtheorem{corollary}{Corollary}

\newtheorem{example}{Example}
\newtheorem{definition}{Definition}

\graphicspath{{./Images/}}

\title[Image Decomposition]{Image Decomposition: Theory, Numerical Schemes, and Performance Evaluation}
\author{J\'er\^ome Gilles}
\address{DGA/CEP - EORD department,\\ 16bis rue Prieur de la C\^ote d'Or, 94110 Arcueil, France}
\email{jerome.gilles@etca.fr}

\begin{document}

\begin{abstract}
This paper describes the many image decomposition models that allow to separate structures and textures or structures, textures, and noise. These models combined a total variation approach with different adapted functional spaces such as Besov or Contourlet spaces or a special oscillating function space based on the work of Yves Meyer. We propose a method to evaluate the performance of such algorithms to enhance understanding of the behavior of these models.\\
\end{abstract}

\maketitle


\tableofcontents
\section{Introduction}

In the last few years, different algorithms have been proposed to decompose an image into its structures and textures components and then its structures, textures, and noise components. The initial idea was proposed by \cite{meyer}. He proposed starting from the Rudin-Osher-Fatemi algorithm, (\cite{rof}), which was designed to perform image denoising. Meyer showed that this model rejects the textures and then proposed to use a new function space, $G$, by replacing the $L^2$-norm by the $G$-norm. He proved that this space corresponds to a space of oscillating functions that are useful to model textures. Two years later, two numerical schemes were proposed to solve Meyer's model, particularly the algorithm based on Chambolle's nonlinear projector. It is easy to implement, and convergence conditions are given by a theorem.

These models work well provided no noise is present in the image. Otherwise, it is necessary to extend the model to a three-part model. Different approachs were proposed based on a local adaptable algorithm or wavelet soft thresholding \cite{aujoluvw,jegilles2}.

This paper describes the philosophy developped by Meyer and gives a description of the different structures + textures models in section \ref{sec:2dec} and structures + textures + noise models in section \ref{sec:3dec}. A new three-part model, based on contourlet soft thresholding, is introduce. This mode improves the results of the previous algorithms.

Section \ref{sec:eval} deals with performance evaluation of the decomposition algorithms. A specific methodology is proposed. First we create test images by recomposing structures, textures, and noise reference images that are generated separatly. We define some metrics to evaluate the quality of the different components obtained at the output of the decomposition algorithms (especially, the problem of how to measure the remaining residue in the noise).

Before detailing the different decomposition models, the first section provides some preliminaries and notations like the wavelet, contourlet formalism. It also presents the different function spaces and their associated norms that are used in the remainder of the paper.

We conclude by summarizing the different models and their performance. We also give some perspectives to this work.

% %==============================================================================
% %  PRELIMINARIES
% %==============================================================================
\section{Preliminaries}
This section describes all the definitions used in the chapter. We start by recalling the multiresolution formalism, specially based on wavelets and other geometric approachs like curvelets and contourlets. We also introduce different function spaces like the space of bounded variations functions ($BV$), Besov spaces, and so on. We complete these descriptions by defining a space based on the contourlet expansion, which will be used in the new three-part decomposition model presented in section \ref{sec:contdec}.

\subsection{Wavelets}

Let us start with the notations and properties of wavelet analysis. The first wavelet expansion of a one dimensional (1D) signal appeared in the 1980's (\cite{berlin,mallat,wkids}). The well-known contributors of the wavelet theory are, but not restricted to, \cite{meyer3}, \cite{mallat}, and \cite{daubechies}.\\

In the following, we assume that we have a 1D signal but the $D$-dimensional extension is naturally obtained by using $D$ separable transforms along the different variables. Wavelet analysis outperforms the Fourier representation. Fourier transform decomposes a signal over a sine-cosine basis. This transform is well localized in frequency but not in time (sine and cosine functions are defined over an infinite domain). For example, if we analyze a transient phenomenon, its Fourier transform covers all the frequency plane while it is well localized in time. It is evident that a transform that is both localized in time and frequency is needed. The first solution used a windowed-Fourier transform. It allows decomposition of the time-frequency plane into many time-frequency atoms. However, this transform is not completely satisfactory because it does not authorize adaptable atoms. However we could be interested in analyzing many transient phenomena with different lengths, then adaptable atoms are necessary. The wavelet transform affords us this opportunity, and we now recall its definition.\\

\subsubsection{Continuous Case}
Wavelet transform decomposes a signal over a set of translated and dilated versions of a mother wavelet. A mother wavelet is a function $\psi\in L^2(\R)$ that respects some criteria as follows:

\begin{equation}
    \int_{\R}\psi(t)dt=0 \qquad \text{zero mean,}
\end{equation}

\begin{equation}
    \|\psi\|_{L^2} = 1 \qquad \text{normalized,}
\end{equation}
and $\psi$ needs to be centered on $0$. If we denote $a$ and $b$ as the dilation and translation parameters, respectively, then the set of wavelets is obtained from the mother wavelet $\psi$ by

\begin{equation}
    \psi_{a,b}(t)=\frac{1}{\sqrt{a}}\psi\left(\frac{t-b}{a}\right).
\end{equation}

Then, we can define the wavelet transform of a function $f\in L^2(\R)$ at time $b$ and scale $a$ by ($\psi^*$ is the complex conjugate of $\psi$)

\begin{equation}
    \mathcal{WT}_f(a,b)=\langle f,\psi_{a,b} \rangle = \int_{\R}f(t)\frac{1}{\sqrt{a}}\psi^*\left(\frac{t-b}{a}\right)dt.
\end{equation}

It is easy to see that a wavelet transform can be written as a convolution product (denoted $\star$)
\begin{equation}
    \mathcal{WT}_f(a,b)=f\star \bar{\psi}_a(b), \qquad \text{where} \quad \bar{\psi}_a(t)=\frac{1}{\sqrt{a}}\psi^*\left(\frac{-t}{a}\right).
\end{equation}

The following theorem gives the conditions that permit reconstruction of the function $f$ from its wavelet expansion.

\begin{thm}
    Let $\psi\in L^2(\R)$ be a real wavelet that respects the following admissibility condition:
    \begin{equation}
        C_{\psi}=\int_0^{+\infty}\frac{|\hat{\psi}(\xi)|^2}{\xi}d\xi < +\infty,
    \end{equation}
    where $\hat{\psi}$ is the Fourier transform of $\psi$. Then, all functions $f\in L^2(\R)$ verify
    \begin{equation}
        f(t)=\frac{1}{C_{\psi}}\int_0^{+\infty}\int_{\R}\mathcal{WT}_f(a,b)\frac{1}{\sqrt{a}}\psi\left(\frac{t-b}{a}\right)db\frac{da}{a^2}
    \end{equation}
    and (Parseval relation)
    \begin{equation}
        \int_{\R}|f(t)|^2dt=\frac{1}{C_{\psi}}\int_0^{+\infty}\int_{\R}|\mathcal{WT}_f(a,b)|db\frac{da}{a^2}.
    \end{equation}
\end{thm}

A proof can be find in \cite{mallat}.\\

Many papers in the literature deal with the choice of the mother wavelet
$\psi$. According to the concerned applications, we can impose some complementary constraints to the wavelet (e.g. its regularity, the length of its support, the number of its zero moments).

\subsubsection{Discrete Case}
In practice, we have digital signals composed of $N$ samples denoted $f[n]$. Let $\psi (t)$ be a continous wavelet where its support is $[-K/2$, $K/2]$; then the discrete wavelet, dilated by $2^j$, is defined as

\begin{equation}
    \psi_{jn}[k]=\frac{1}{\sqrt{2^j}}\psi[2^{-j}k-n].
\end{equation}
Then the discrete wavelet transform can be written as

\begin{equation}
    \mathcal{WT}_f[n,j]=\sum_m f[m]\psi_{jn}^*[m]=\langle f,\psi_{jn} \rangle,
\end{equation}
and the reconstruction formula is true if $\psi$ has some complementary properties, (see \cite{mallat} for more details). Then, we have

\begin{equation}
    f[m]=\sum_{j=0}^{+\infty}\sum_n \mathcal{WT}_f[n,j]\psi_{jn}[n].
\end{equation}

These relations show that filter banks, defined from $\psi$, can be used to implement the wavelet transform and its inverse.

\subsection{Multiresolution Analysis}\label{sec:wavelets}
Multiresolution analysis is defined in \cite{mallat}. Let $\{V_j\}_{j\in\Z}$ be a set of closed subspaces of $L^2(\R)$. We said it is a multiresolution approximation if it meets the following conditions:

\begin{gather}
    \forall (j,k)\in\Z^2 \; ,\; f(t)\in V_j \Leftrightarrow f(t-2^jk) \in V_j ,\\
    \forall j\in\Z \; , \; V_{j+1} \subset V_j ,\\
    \forall j\in\Z \; , \; f(t)\in V_j \Leftrightarrow f\left(\frac{t}{2}\right)\in V_{j+1} ,\\
    \lim_{j\rightarrow +\infty} V_j = \bigcap_{j=-\infty}^{+\infty} V_j=\{0\} ,\\ \notag
    \\
    \lim_{j\rightarrow -\infty} V_j = \overline{\bigcup_{j=-\infty}^{+\infty} V_j}=L^2(\R),
\end{gather}
and there exists a function $\theta$ such that $\{\theta (t-n)\}_{n\in\Z}$ is a Riesz basis of $V_0$.\\

Let $\varphi$ be a function (called the scale function) with its Fourier transform be defined by:
\begin{equation}
    \hat{\varphi}(\omega)=\frac{\hat{\theta}(\omega)}{\left(\sum_{k=-\infty}^{+\infty}|\hat{\theta}(\omega+2k\pi)|^2\right)^{1/2}}.
\end{equation}

Then the set $\{\varphi_{jn}\}_{n\in\Z}$ defined by
\begin{equation}
    \varphi_{jn}(t)=\frac{1}{\sqrt{2^j}}\varphi\left(\frac{t-n}{2^j}\right)
\end{equation}
is an orthonormal basis of $V_j$. If we define $W_j=V_j \ominus V_{j+1}$, the wavelet set $\{\psi_{jn}\}_{n\in\Z}$ associated with $\varphi$ (see \cite{berlin,mallat,wkids} to learn to build such functions) is an orthonormal basis of $W_j$. Then all functions $f\in L^2(\R)$ can be decomposed to

\begin{equation}\label{eq:decompw}
    f(t)=\sum_n \alpha_n\varphi_{0n}(t)+\sum_{j=0}^{+\infty}\sum_n\beta_{jn}\psi_{jn}(t),
\end{equation}
where the coefficients $\beta_{jn}=\langle f,\psi_{jn}\rangle$ are the wavelet transform coefficients and $\alpha_{n}=\langle f,\varphi_{0n}\rangle$ are the coefficients from the projection on the subspace $V_0$. In other terms, we have

\begin{equation}
    \text{(\ref{eq:decompw})} \Longleftrightarrow f\in V_0 \oplus \bigoplus_{j=0}^{\infty}W_j.
\end{equation}

\subsection{Directional Multiresolution Analysis}
The two dimensional (2D) extension of a wavelet generally uses the separability principle. It uses a 1D wavelet filter along the horizontal and vertical directions. In natural images, however, the information is not limited to these two directions. It is easy to understand that the multiresolution analysis needs to be extended to encompass directions in the image. Many authors propose different approachs to do this directional analysis. This chapter describes only those best known in the literature: the ridgelets, curvelets, and contourlets.

\subsection{Ridgelets}
In his doctoral dissertation, \cite{candes} proposes a new transform that deals with directionality in images: the ridgelet transform.

The ridgelets functions $\psi_{a,b,\theta}$ are defined in a manner similar to wavelets but add the notion of orientation (tuned by the $\theta$ parameter):
\begin{align}
    \psi_{a,b,\theta}           & : \R^2 \longrightarrow \R^2                                                  \\
    \psi_{a,b,\theta}(x_1,x_2)= & \frac{1}{\sqrt{a}}\psi \left(\frac{x_1\cos\theta+x_2\sin\theta-b}{a}\right).
\end{align}

The $\psi_{a,b,\theta}$ is constant along the lines $x_1\cos\theta+x_2\sin\theta=c$ ($c$ is a constant) and is a wavelet $\psi$ in the orthogonal direction. Many properties of the wavelet theory can be transposed.\\

\begin{definition}
    The admissibility condition for a ridgelet is:
    \begin{equation}
        K_{\psi}=\int \frac{\left|\hat{\psi}(\xi)\right|^2}{\left|\xi\right|^2}d\xi <\infty,
    \end{equation}
    which is equivalent to $\int\psi(t)dt=0$.
\end{definition}

Morever, we assume that $\psi$ is normalized:
\begin{equation}
    \Rightarrow \int \frac{\left|\hat{\psi}(\xi)\right|^2}{\left|\xi\right|^2}d\xi=1.
\end{equation}

Under these assumptions, Cand\`es defines the ridgelet transform of a function $f$ by

\begin{definition}
    For a function $f$, the coefficients of its ridgelet transform are given by
    \begin{equation}\label{eq:ridgelet}
        \mathcal{R}_f(a,b,\theta)=\int\psi_{a,b,\theta}^*(x_1,x_2)f(x_1,x_2)dx_1dx_2=<f,\psi_{a,b,\theta}>,
    \end{equation}
    and the reconstruction formula is given by
    \begin{equation}
        f(x_1,x_2)=\int_{0}^{2\pi}\int_{-\infty}^{+\infty}\int_{0}^{+\infty}\mathcal{R}_f(a,b,\theta)\psi_{a,b,\theta}(x)\frac{da}{a^3}db\frac{d\theta}{4\pi}.
    \end{equation}
\end{definition}

In addition, the Parseval relation is verified as in proposition \ref{prop:parseval} below,\\
\begin{proposition}\label{prop:parseval}
    If $f\in L^1\cap L^2(\R^2)$ and if $\psi$ is admissible, then
    \begin{equation}
        \|f\|_{L^2}^2=c_{\psi}\int |\langle f,\psi_{a,b,\theta}\rangle |^2\frac{da}{a^3}db\frac{d\theta}{4\pi},
    \end{equation}
    where $c_{\psi}=(4\pi)^{-1}K_{\psi}^{-1}$.
\end{proposition}
The proof can be found in \cite{candes}.

In pratice, the ridgelet transform can be implemented by using the Radon transform and the 1D wavelet transform (see \cite{candes} for more details).

\subsection{Curvelets}
From the definition of the ridgelet transform, it is easy to see that this transform is a global transform (we mean that it is efficient to represent lines that go through the entire image). But images contain more general edges that are present locally. \cite{candes99curvelets,candesfdct,donoho99digital} propose a new approach that provides a local directional multiresolution analysis called the curvelet transform.\\
The idea is to do a specific tiling of the space and frequency planes by using two windows, the radial window $W(r)$ and the angular window $V(t)$, where $(r,\theta)$ are the polar coordinates in the frequency plane and $r\in (1/2,2)$. The window $V$ is defined for $t\in [-1,1]$. These windows obey the following admissibility conditions:
\begin{equation}
    \sum_{j=-\infty}^{+\infty}W^2(2^jr)=1 \qquad r\in(3/4,3/2)
\end{equation}
and
\begin{equation}
    \sum_{l=-\infty}^{+\infty}V^2(t-l)=1 \qquad t\in(-1/2,1/2)
\end{equation}
Then for each $j\geqslant j_0$, a frequency window $U$ is defined in the Fourier domain by
\begin{equation}
    U_j(r,\theta)=2^{-3j/4}W(2^{-j}r)V\left(\frac{2^{\lfloor j/2\rfloor}\theta}{2\pi}\right),
\end{equation}
where $\lfloor j/2\rfloor$ is the integer part of $j/2$. Let $\varphi_j(x)$ denote the function such that its Fourier transform $\hat{\varphi}_j(\omega)=U_j(w)$ ($(r,\theta)$ are the polar coordinates corresponding to $w=(w_1,w_2)$). Then we define at scale $2^{-j}$, orientation $\theta_l$, and position $x_k^{(j,l)}$ a set of curvelets by
\begin{equation}
    \varphi_{j,l,k}(x)=\varphi_j\left(R_{\theta_l}(x-x_k^{(j,l)}) \right),
\end{equation}
where $R_{\theta_l}$ is the rotation by $\theta_l$ radians. Then the curvelet transform is simply defined by the inner product between a function $f\in \R^2$ with the set of curvelets. A curvelet coefficient can be written
\begin{equation}
    c(j,l,k)=\langle f,\varphi_{j,l,k} \rangle=\int_{\R^2}f(x)\varphi_{j,l,k}^*(x)dx.
\end{equation}
More details can be found in \cite{candesfdct}. In their paper, the authors prove the following proposition.

\begin{proposition}
    Let $f\in L^2(\R^2)$ denote a function expanded over a set of curvelets $\varphi_{j,l,k}$; we have the following reconstruction formula:
    \begin{equation}
        f=\sum_{j,l,k}\langle f,\varphi_{j,l,k} \rangle \varphi_{j,l,k} \qquad \text{(Tight frame,)}
    \end{equation}
    and the Parseval relation is verified:
    \begin{equation}
        \sum_{j,l,k}\left| \langle f,\varphi_{j,l,k} \rangle \right|^2=\|f\|_{L^2}^2, \qquad \forall f\in L^2(\R^2).
    \end{equation}
\end{proposition}

All details about the numerical aspects can be found in \cite{candesfdct}.

\subsection{Contourlets}\label{sec:contourlet}
\begin{figure}[ht]
    \psfrag{f(x)}{f}
    \psfrag{LP}{\footnotesize{LP}}
    \psfrag{DFB1}{\footnotesize{DFB1}}
    \psfrag{DFB2}{\footnotesize{DFB2}}
    %\centering\includegraphics[scale=0.5]{contourlet.eps}
    \centering\includegraphics[scale=1.2]{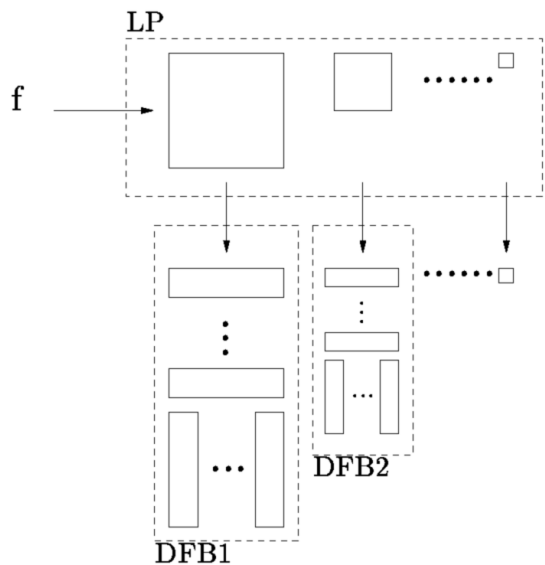}
    \caption{Contourlet transform principle.}
    \label{fig:pdfb}
\end{figure}

In 1999, when Cand\`es et al. proposed the curvelet transform, the authors showed many promising results. The main drawback of the first version of curvelets is the difficulty of its numerical implementation (the discrete curvelet transform was proposed in 2005 \cite{candesfdct}). In order to ``overcome'' this problem, \cite{do7,do4,do6,do5,do1,do3,do2}  proposed a new algorithm, called the contourlet transform, initially designed in a discrete framework. The idea is to combine a multiscale decomposition and directional filtering at each scale (Figure \ref{fig:pdfb}).

The multiscale decomposition is obtained by using a Laplacian pyramid decomposition (LP) (\cite{burt}). The directional filtering uses a directional filter bank (DFB) based on quincunx filters (\cite{bamberger}). In the next theorem, the authors show that this transform produces a tight frame.

\begin{thm}
    Let $j$ be the scale, $n$ the position, $\left\{l_j\right\}_{j\leqslant j_0}$ the set of number of directions for each scale $j$. Then, the set
    \begin{equation}
        \left\{\phi_{j_0,n}(t);\rho^{(l_j)}_{j,k,n}(t)\right\}_{j\leqslant j_0, \; 0\leqslant k\leqslant 2^{l_j}-1, \;n\in\Z^2}
    \end{equation}
    is a tight frame of $L_2(\R^2)$.
\end{thm}

All details about the construction of functions $\phi_{j_0,n}(t)$ and $\rho^{(l_j)}_{j,k,n}(t)$ can be found in \cite{do7}.

This implies
\begin{corollary}\label{cor:contourlet}
    \begin{equation}
        f(t)=\sum_n \alpha_n \phi_{j_0,n}(t)+\sum_{j\leqslant j_0}\sum_{k=0}^{2^{l_j}-1}\sum_n \beta_{j,k,n}\rho^{(l_j)}_{j,k,n}(t)
    \end{equation}
    or
    \begin{equation}
        f(t)=\sum_{j\in \Z}\sum_{k=0}^{2^{l_j}-1}\sum_n \beta_{j,k,n}\rho^{(l_j)}_{j,k,n}(t),
    \end{equation}
\end{corollary}
where $\alpha_n=\langle f|\phi_{j_0,n}\rangle$ and $\beta_{j,k,n}=\langle f|\rho_{j,k,n}^{(l_j)}\rangle$ are the contourlet transform coefficients.

\subsection{Function Spaces}
In Sections \ref{sec:2dec} and \ref{sec:3dec}, we will use some function spaces and more particularly their associated norms. This section briefly describes the spaces of interest (it is assumed that the reader knows the $L^p$ spaces and $d$ is the dimension). The goal of the different spaces is to characterize some properties like the differentiability and the regularity of functions.

\subsubsection{Sobolev Spaces}
The first spaces we are interested in are the Sobolev spaces $W^{k,p}$. These spaces are defined as the spaces of functions $f$ such that they and their weak derivatives up to some order $k$ have a finite $L^p$ norm, for a given $p\geqslant 1$. These spaces are endowed with the following norm:
\begin{equation}
    \|f\|_{W^{k,p}}=\left(\sum_{i=0}^k \|f^{(i)}\|_{L^p}^p\right)^{1/p}=\left(\sum_{i=0}^k \int |f^{(i)}(t)|^pdt\right)^{1/p}.
\end{equation}

An interesting particular case is for $p=2$, denoted $H^k=W^{k,2}$, because of their relation with the Fourier series. More information about the Sobolev spaces can be found in the book by \cite{adams}.

\subsubsection{Besov Spaces}
The next kind of spaces are Besov spaces $B_{p,q}^s$. Functions taken in $B_{p,q}^s$ have $s$ derivatives in $L^p$. The parameter $q$ permits more precise characterization of the regularity. A general description of these spaces can be found in \cite{triebel2}. In this paper, we give only their connection with wavelets. Indeed, different expressions exists for the norm associated with Besov space but one uses the wavelet coefficients, see (\ref{eq:defbesov}).
\begin{align}\label{eq:defbesov}
    \forall f\in B_{p,q}^s \qquad \|f\|_{B_{p,q}^s} & =\left[\sum_n |\alpha_n|^p\right]^{1/p} \notag  \\
    & +\left(\sum_{j=0}^{+\infty}2^{j\left(\frac{d}{2}-\frac{1}{p}+s\right)q}\left[\sum_n 2^{j\frac{p}{2}}|\beta_{jn}|^p\right]^{q/p}\right)^{1/q}.
\end{align}

The homogeneous version is
\begin{equation}\label{eq:hdefbesov}
    \forall f\in \dot{B}_{p,q}^s \qquad \|f\|_{\dot{B}_{p,q}^s}=\left(\sum_{j=-\infty}^{+\infty}2^{j\left(\frac{d}{2}-\frac{1}{p}+s\right)q}\left[\sum_n 2^{j\frac{p}{2}}|\beta_{jn}|^p\right]^{q/p}\right)^{1/q},
\end{equation}
where $\alpha_n$ and $\beta_{jn}$ are the coefficients issued from the wavelet expansion (see Section \ref{sec:wavelets}).

\subsubsection{Ridgelet Spaces}
In the same way as previous, Cand\`es define the ridgelet spaces $R_{p,q}^s$ endowed with the norm based on the ridgelet coefficients.
\begin{definition}
    For $s\geqslant 0$ and $p,q>0$, we said that $f\in R_{p,q}^s$ if $f\in L^1$ and
    \begin{multline}
        \underset{u}{Ave} \|R_f(u,.)\star\varphi\|_{L^p}<\infty \\
        \text{and} \quad \left\{2^{js}2^{j(d-1)/2}\left(\underset{u}{Ave} \|R_f(u,.)\star\psi_j\|_{L^p}^p\right)^{1/p}\right\}\in l_q(\N),
    \end{multline}
    where $R_f(u,t)=\int_{u.x=t}f(x)dx$ is the Radon transform of $f$ ($u=(\cos\theta ; \sin\theta)$). The function $\varphi$ is the scale function associated with $\psi$.
\end{definition}

Then the induced norm is defined by
\begin{multline}
    \|f\|_{R_{p,q}^s}=\underset{u}{Ave} \|R_f(u,.)\star\varphi\|_{L^p} \\
    +\left\{\sum_{j \geqslant 0}\left(2^{js}2^{j(d-1)/2}\left(Ave_u \|R_f(u,.)\star\psi_j\|_{L^p}^p\right)^{1/p}\right)^q\right\}^{1/q}
\end{multline}
and its homogeneous version $\dot{R}_{p,q}^s$
\begin{equation}
    \|f\|_{\dot{R}_{p,q}^s}=\left\{\sum_{j \in \Z}\left(2^{js}2^{j(d-1)/2}\left(Ave_u \|R_f(u,.)\star\psi_j\|_{L^p}^p\right)^{1/p}\right)^q\right\}^{1/q}.
\end{equation}
As in the Besov case, these norms can be calculated from the ridgelet coefficients. Let $w_j(u,b)(f)=\langle f(x),\psi_j(u.x-b)\rangle$ for $j\geqslant 0$ and $v(u,b)(f)=\langle f(x),\varphi(u.x-b)\rangle$ these  ridgelet coefficients, then
\begin{multline}
    \|f\|_{R_{p,q}^s}=\left(\int |v(u,b)(f)|^pdudb\right)^{1/p}\\
    +\left\{\sum_{j\geqslant 0}\left(2^{js}2^{j(d-1)/2}\left(\int |w_j(u,b)(f)|^pdudb\right)^{1/p}\right)^q\right\}^{1/q}.
\end{multline}

More information can be found in \cite{candes}.

\subsubsection{Contourlet Spaces}\label{sec:contspace}
Inspired from the previous spaces, we propose to define the contourlet spaces, which will be denoted $Co_{p,q}^s$.
\begin{definition}
    Let $s\geqslant 0$ and $p,q>0$, if $f\in Co_{p,q}^s$; then

    \begin{align} \label{eq:defctspq}
        \|f\|_{Co_{p,q}^s} & =\left[\sum_n |\alpha_{j_0,n}|^p \right]^{1/p} \notag                                                                                                                     \\
                           & +\left\{\sum_{j\leqslant j_0}2^{j\left(\frac{d}{2}-\frac{1}{p}+s\right)q}\left[\sum_{k=0}^{2^{l_j}-1}\sum_n 2^{j\frac{p}{2}}|\beta_{j,k,n}|^p\right]^{q/p}\right\}^{1/q},
    \end{align}
    or in the homogeneous case,
    \begin{equation} \label{eq:defctspq2}
        \|f\|_{\dot{Co}_{p,q}^s}=\left\{\sum_{j\in \Z}2^{j\left(\frac{d}{2}-\frac{1}{p}+s\right)q}\left[\sum_{k=0}^{2^{l_j}-1}\sum_n 2^{j\frac{p}{2}}|\beta_{j,k,n}|^p\right]^{q/p}\right\}^{1/q},
    \end{equation}
    where $\alpha_{j_0,n}$ and $\beta_{j,k,n}$ are the contourlet coefficients mentioned in Section \ref{sec:contourlet}.
\end{definition}

\subsubsection{Bounded Variation ($BV$) Spaces}
The last space of interest is the $BV$ space, the space of bounded variations functions. This space is widely used in image processing because it is a good candidate to modelize structures in images.
\begin{definition}
    The space $BV$ over a domain $\Omega$ is defined as
    \begin{equation}
        BV=\left\lbrace f\in L^1(\Omega); \int_{\Omega} |\nabla f|<\infty \right\rbrace,
    \end{equation}
    where $\nabla f$ is the gradient, in the distributional sense, of $f$ and
    \begin{equation}
        \int_{\Omega} |\nabla f|=\underset{\overrightarrow{\varphi}}{\sup}\left\lbrace \int_{\Omega} f \Div \overrightarrow{\varphi}; \qquad \overrightarrow{\varphi}\in C_0^1(\Omega,\R^2),|\overrightarrow{\varphi}|\leqslant 1 \right\rbrace.
    \end{equation}
\end{definition}
This space is endowed with the following norm:
\begin{equation}
    \|f\|_{BV}=\|f\|_{L^1}+\int_{\Omega}|\nabla f|.
\end{equation}
But in general, we only keep the second term, which is well known as the total variation of $f$. In the rest of the paper, we will use the notation
\begin{equation}
    J(f)=\int_{\Omega}|\nabla f|.
\end{equation}
More information about the $BV$ space is available in \cite{haddad,vesephd}.\\

We now have all the basic tools needed to describe the image decomposition models. The next two sections present the structures $+$ textures and structures $+$ textures $+$ noise models, respectively.

% %==============================================================================
% %  STRUCTURES + TEXTURES
% %==============================================================================
\section{Structures + Textures Decomposition}\label{sec:2dec}

The starting point of the image decomposition models is the work of \cite{meyer} about the Rudin-Osher-Fatemi (ROF) algorithm (\cite{rof}). Let us recall the ROF model. Assume $f$ is an observed image that is the addition of the ideal scene image $u$, which we want to retrieve, and a noise $b$. The authors propose to minimize the following functional to get $u$:

\begin{equation}\label{eq:rof2bis}
    F_{\lambda}^{ROF}(u)=J(u)+\lambda \|f-u\|_{L^2}^2.
\end{equation}

This model assumes that $u$ is in $BV$ because this space preserves sharp edges. This algorithm gives good results and is very easy to implement by using the nonlinear projectors proposed by \cite{chambolle} (see Appendix \ref{ap:chambolle}).

Now if we take the image decomposition point of view, $f=u+v$, the functional in Eq.(\ref{eq:rof2bis}) can be rewritten as
\begin{equation}\label{eq:rof2mod}
    F_{\lambda}^{ROF}(u,v)=J(u)+\lambda \|v\|_{L^2}^2.
\end{equation}

We remind the reader that decomposition means $u$ is the structures part and $v$ the textures part.
Meyer shows that this model is not adapted to achieve this decomposition. In order to convince us, the following example illustrates that the more a texture is oscillating, the more it is removed from both the $u$ and $v$ parts.

\begin{example}\label{ex:osctexture}
    Let $v$ be a texture created from an oscillating signal over a finite domain. Then $v$ can be written ($x=(x_1,x_2)$) as follows:
    \begin{equation}\label{eq:osctexture}
        v(x)=\cos(\omega x_1)\theta(x),
    \end{equation}
    where $\omega$ is the frequency and $\theta$ the indicator function over the considered domain. Then we can calculate the $L^2$ and $BV$ norms of $v$, respectively. We get
    \begin{equation}
        \|v\|_{L^2}\approx \frac{1}{\sqrt{2}}\|\theta\|_{L^2},
    \end{equation}
    which is constant $\forall \omega$ and does not specially capture textures. In addition,
    \begin{equation}
        \|v\|_{BV}=\frac{\omega}{2\pi}\|\theta\|_{L^1},
    \end{equation}
    which grows as $\omega \rightarrow \infty$ and then clearly rejects textures.
\end{example}

In order to adapt the ROF model to capture the textures in the $v$ component, Meyer proposes to replace $L^2$ space by another space, called $G$, which is a space of oscillating functions. He proves that this space is the dual space of $\mathcal{BV}$ (where $\mathcal{BV}=\{f\in L^2(\R^2)\; , \; \nabla f\in L^1(\R^2)\}$, which is close to the $BV$ space and the total variation described earlier in the paper); see \cite{meyer} for more theoretical details about these spaces.

This space $G$ is endowed by the following norm:
\begin{equation}\label{eq:normeg}
    \|v\|_G=\inf_{g}\left\|\left(\left|g_1\right|^2+\left|g_2\right|^2\right)^{\frac{1}{2}}\right\|_{L^{\infty}},
\end{equation}
where $g=(g_1,g_2)\in L^{\infty}(\R^2)\times L^{\infty}(R^2)$ and $v=\Div g$. If we calculate the $G$-norm of the oscillating texture in Eq.(\ref{eq:osctexture}) of example \ref{ex:osctexture}, we get
\begin{equation}\label{eq:normegcos}
    \|v\|_G \leqslant \frac{C}{|\omega|},
\end{equation}
where $C$ is a constant. Then it is easy to see that this space $G$ is well adapted to capture textures. Now, the modified functional performing the structures $+$ textures decomposition is

\begin{equation}\label{eq:modmeyer}
    F_{\lambda}^{YM}(u,v)=J(u)+\lambda\|v\|_G,
\end{equation}
where $f=u+v$, $f\in G$, $u \in BV$, $v\in G$. The drawback of this model is the presence of an $L^{\infty}$ norm in the the expression of the $G$-norm (this does not allow classic variational calculus).

The first people who proposed a numerical algorithm to solve the Meyer model were \cite{vese1}. Their approach was to use the theorem which tells that $\forall f \in L^{\infty}(\Omega), \|f\|_{L^{\infty}}=\lim_{p\rightarrow \infty} \|f\|_{L^p}$ and a slightly modified version of Meyer's functional:
\begin{equation}
    F_{\lambda,\mu,p}^{OV}(u,g)=J(u)+\lambda \|f-(u+\Div \; g)\|_{L^2}^2+\mu \left\|\sqrt{g_1^2+g_2^2}\right\|_{L^p}.
\end{equation}
Then variational calculus applies and results in a system of three connected partial differential equations. All the details of the equations discretization are available in \cite{vese1}. This algorithm works well but is very sensitive in the choice of its parameters, which induced many instability.

Another way to solve Meyer model was proposed by \cite{aujolphd,aujol,aujol2}. The authors propose a dual-method approach that naturally arises because of the dual relation between the $G$ and $BV$ spaces. The problem is assumed to be in the discrete case and defined over a finite domain $\Omega$. They proposed a modified functional to minimize.

\begin{equation}\label{eq:aujolg}
    F_{\lambda,\mu}^{AU}(u,v)=J(u)+J^*\left(\frac{v}{\mu}\right)+(2\lambda)^{-1}\|f-u-v\|_{L^2}^2
\end{equation}

and
\begin{equation}
    (u,v)\in BV(\Omega)\times G_{\mu}(\Omega).
\end{equation}

The set $G_{\mu}$ is the subset in $G$ where $\forall v\in G_{\mu}$, $\|v\|_G\leqslant\mu$. Moreover, $J^*$ is the characteristic function over $G_1$ with the property that $J^*$ is the dual operator of $J$ ($J^{**}=J$). Thus,

\begin{equation}\label{eq:jdual}
    J^*(v)=
    \begin{cases}
        0 \qquad \text{if} \; v\in G_1 \\
        +\infty \qquad \text{else.}
    \end{cases}
\end{equation}

The interesting point is that the precited Chambolle's projectors are the projector over the sets $G_{\mu}, \forall \mu$; these operators will be denoted $P_{G_{\mu}}$ in the rest of the paper. More details about these projectors can be found in \cite{chambolle} and recalled in Appendix \ref{ap:chambolle}. Then the authors propose an iterative algorithm that gives the minimizers $(\hat{u},\hat{v})$ of $F_{\lambda,\mu}^{AU}(u,v)$.

\begin{itemize}
    \item Let us fix $v$, we seek for the minimizer $u$ of
          \begin{equation}
              \label{equ:cha1}
              \inf_{u}\left(J(u)+(2\lambda)^{-1}\|f-u-v\|_{L^2}^2\right).
          \end{equation}
    \item Now we fix $u$ and seek for the minimizer $v$ of
          \begin{equation}
              \label{equ:cha2}
              \inf_{v} J^*\left(\frac{v}{\mu}\right)+\|f-u-v\|_{L^2}^2.
          \end{equation}
\end{itemize}

Chambolle's results show that the solution of Eq.(\ref{equ:cha1}) is given by
\begin{equation}
    \hat{u}=f-\hat{v}-P_{G_{\lambda}}(f-\hat{v})
\end{equation}
and the solution of Eq. (\ref{equ:cha2}) by
\begin{equation}
    \hat{v}=P_{G_{\mu}}(f-\hat{u}).
\end{equation}

Then the numerical algorithm is\\
\begin{center}
    %\fbox{\parbox[h]{0.9\textwidth}{
    \begin{enumerate}
        \item Initialization:
              \begin{equation*}
                  u_0=v_0=0
              \end{equation*}
        \item Iteration $n+1$:
              \begin{eqnarray*}
                  &v_{n+1}=P_{G_{\mu}}(f-u_n)\\
                  &u_{n+1}=f-v_{n+1}-P_{G_{\lambda}}(f-v_{n+1})
              \end{eqnarray*}
        \item We stop the algorithm if
              \begin{equation*}
                  \max\left(|u_{n+1}-u_n|,|v_{n+1}-v_n|\right)\leqslant \epsilon
              \end{equation*}
              or if we reach a prescribed maximal number of iterations.\\
    \end{enumerate}%}}
\end{center}

The authors prove that the minimizers $(\hat{u},\hat{v})$ are also minimizers of the original Meyer functional Eq. (\ref{eq:modmeyer}), and that it is better to start by calculating $v_{n+1}$ than $u_{n+1}$. See \cite{aujolphd,aujol} for the complete proofs.\\

\begin{figure}[!ht]
    \centering
    \begin{tabular}{m{5cm}m{5cm}}
        \includegraphics[width=0.4\textwidth]{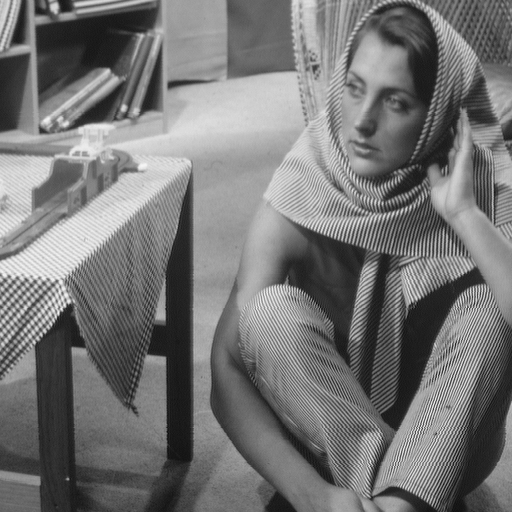} & \includegraphics[width=0.4\textwidth]{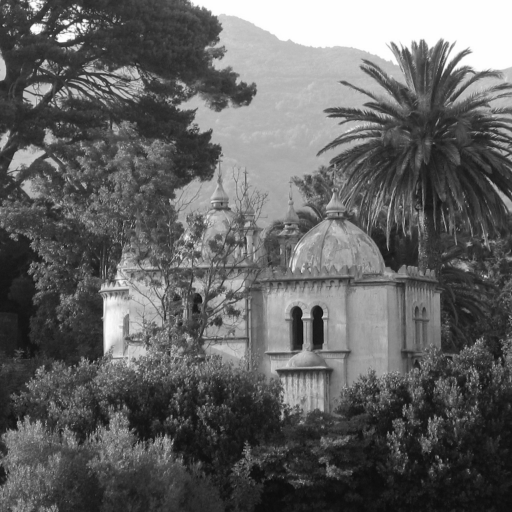} \\
        \includegraphics[width=0.4\textwidth]{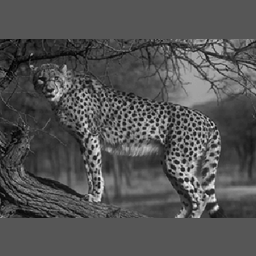} &
    \end{tabular}
    \caption{Original Barbara, House, and Leopard images.}
    \label{fig:imorig}
\end{figure}

\begin{figure}[!ht]
    \centering
    %\begin{tabular}{m{5cm}m{5cm}}
    \begin{tabular}{cc}
        \includegraphics[width=0.4\textwidth]{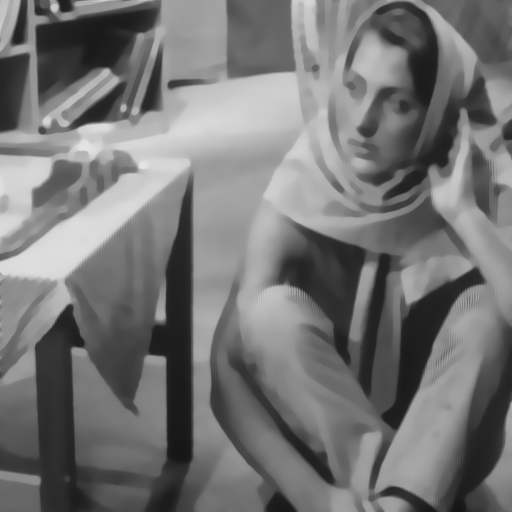} & \includegraphics[width=0.4\textwidth]{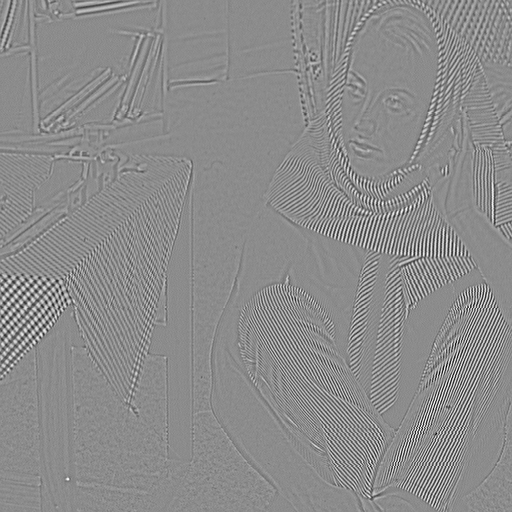} \\
        Structures                                               & Textures
    \end{tabular}
    \caption{$BV$-$G$ structures $+$ textures image decomposition of Barbara image.}
    \label{fig:uvbarb}
\end{figure}

\begin{figure}[!ht]
    \centering
    %\begin{tabular}{m{5cm}m{5cm}}
    \begin{tabular}{cc}
        \includegraphics[width=0.4\textwidth]{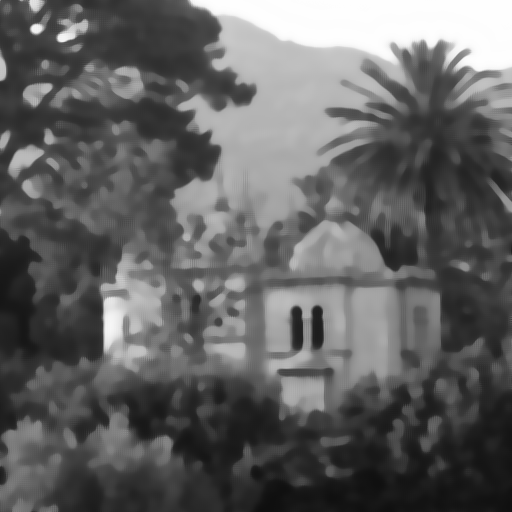} & \includegraphics[width=0.4\textwidth]{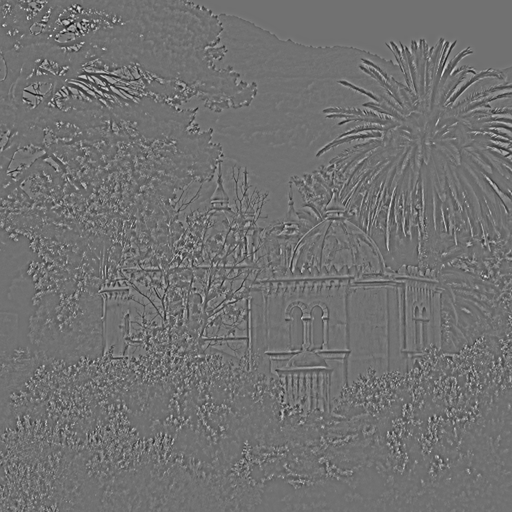} \\
        Structures                                               & Textures
    \end{tabular}
    \caption{$BV$-$G$ structures $+$ textures image decomposition of House image.}
    \label{fig:uvhouse}
\end{figure}

\begin{figure}[!ht]
    \centering
    %\begin{tabular}{m{5cm}m{5cm}}
    \begin{tabular}{cc}
        \includegraphics[width=0.4\textwidth]{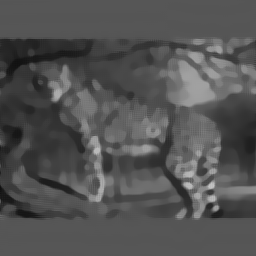} & \includegraphics[width=0.4\textwidth]{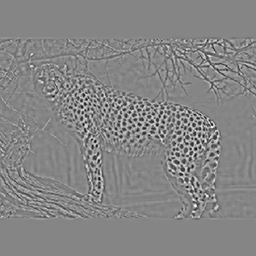} \\
        Structures                                                  & Textures
    \end{tabular}
    \caption{$BV$-$G$ structures $+$ textures image decomposition of Leopard image.}
    \label{fig:uvleo}
\end{figure}

Figure \ref{fig:imorig} presents the three original images (Barbara, House, and Leopard) use for tests in the rest of the paper. Figures \ref{fig:uvbarb}, \ref{fig:uvhouse}, and \ref{fig:uvleo} illustrate the results from Aujol's algorithm. The chosen parameters are $(\lambda=1,\mu=100)$, $(\lambda=10,\mu=1000)$, and $(\lambda=5,\mu=1000)$ respectively. For clarity reasons, we enhanced the contrasts of the textured components. On each test we see that the separation between structures and textures works well. Some residual textures remain in the structures part; this can be explained by the fact the parameter $\lambda$ acts as a tradeoff between the ``power'' of separability and too much regularization of $u$.

As the $G$-norm is difficult to handle, \cite{meyer} proposes to replace the space $G$ by the Besov space $\dot{B}_{-1,\infty}^{\infty}$ because $G\subset \dot{B}_{-1,\infty}^{\infty}$ (in the following, we will denote $E=\dot{B}_{-1,\infty}^{\infty}$). The advantage is that the norm of a function $v$ over this space can be defined from its wavelet coefficients. The corresponding model proposed by Meyer is
\begin{equation}
    F_{\lambda}^{YM2}(u,v)=J(u)+\lambda\|v\|_E
\end{equation}

Aujol and Chambolle were the first to propose a numerical algorithm that uses the space $E$. As previously, they reformulated the model in a dual-method approach, where $E_{\mu}$ is the subset of $E$, where $\forall f\in E_{\mu}$, $\|f\|_E\leqslant \mu$ and $B^*(f)$ is the indicator function over $E_1$. Then the functional to minimize is

\begin{equation}
    F_{\lambda,\mu}^{AC}(u,v)=J(u)+B^*\left(\frac{v}{\mu}\right)+(2\lambda)^{-1}\|f-u-v\|_{L^2}^2.
\end{equation}

\cite{chambolle2} proved the existence of a projector on this space, denoted $P_{E_{\mu}}$, defined by
\begin{equation}
    P_{E_{\mu}}(f)=f-WST(f,2\mu),
\end{equation}
where $WST$ is the wavelet soft thresholding operator (we mean that we first perform the wavelet expansion of the function, then we do the soft thresholding of the wavelet coefficients, and end by reconstructing the image). Then the new numerical algorithm is as follows:

\begin{center}
    %\fbox{\parbox[h]{0.9\textwidth}{
    \begin{enumerate}
        \item Initialization:
              \begin{equation*}
                  u_0=v_0=0
              \end{equation*}
        \item Iteration $n+1$:
              \begin{eqnarray*}
                  &v_{n+1}=P_{E_{\mu}}(f-u_n)=f-u_n-WST(f-u_n,2\mu)\\
                  &u_{n+1}=f-v_{n+1}-P_{G_{\lambda}}(f-v_{n+1})
              \end{eqnarray*}
        \item We stop if
              \begin{equation*}
                  \max\left(|u_{n+1}-u_n|,|v_{n+1}-v_n|\right)\leqslant \epsilon
              \end{equation*}
              or if we reach a prescribed maximal number of iterations.\\
    \end{enumerate}%}}
\end{center}

\begin{figure}[!ht]
    \centering
    %\begin{tabular}{m{5cm}m{5cm}}
    \begin{tabular}{cc}
        \includegraphics[width=0.4\textwidth]{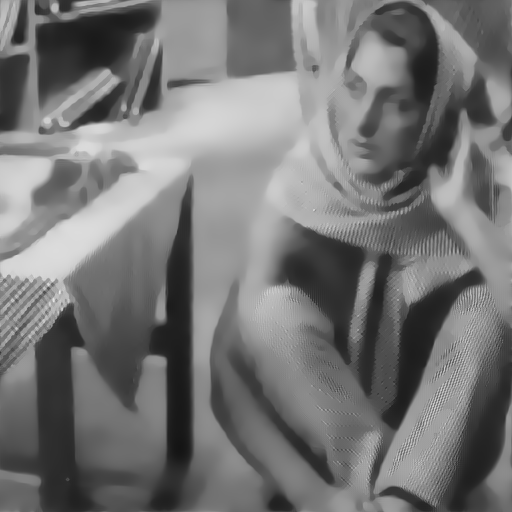} & \includegraphics[width=0.4\textwidth]{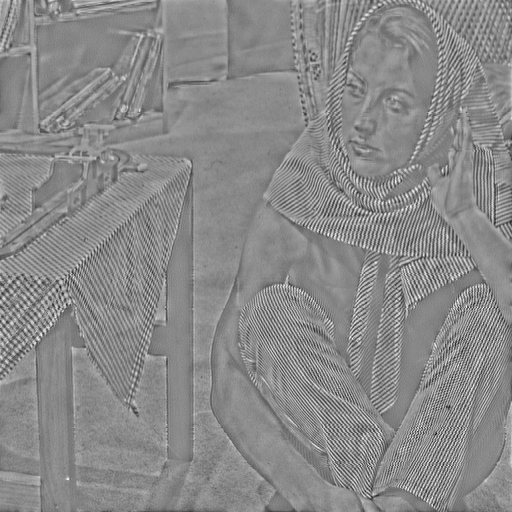} \\
        Structures                                                 & Textures
    \end{tabular}
    \caption{$BV$-$E_{\mu}$ structures $+$ textures image decomposition of Barbara image.}
    \label{fig:uvbbarb}
\end{figure}

\begin{figure}[!ht]
    \centering
    %\begin{tabular}{m{5cm}m{5cm}}
    \begin{tabular}{cc}
        \includegraphics[width=0.4\textwidth]{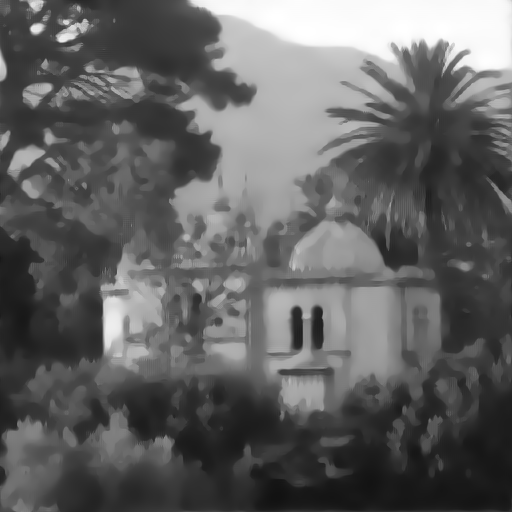} & \includegraphics[width=0.4\textwidth]{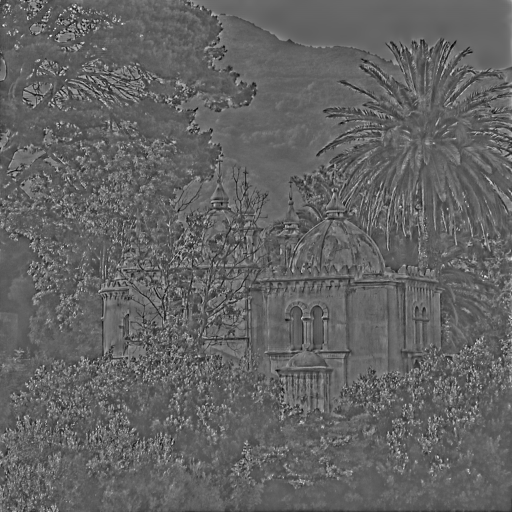} \\
        Structures                                                  & Textures
    \end{tabular}
    \caption{$BV$-$E_{\mu}$ structures $+$ textures image decomposition of House image.}
    \label{fig:uvbhouse}
\end{figure}

\begin{figure}[!ht]
    \centering
    %\begin{tabular}{m{5cm}m{5cm}}
    \begin{tabular}{cc}
        \includegraphics[width=0.4\textwidth]{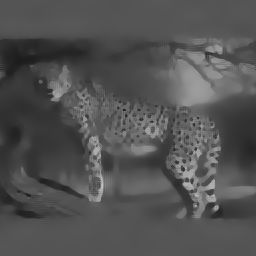} & \includegraphics[width=0.4\textwidth]{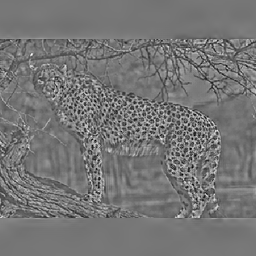} \\
        Structures                                                 & Textures
    \end{tabular}
    \caption{$BV$-$E_{\mu}$ structures $+$ textures image decomposition of Leopard image.}
    \label{fig:uvbleo}
\end{figure}

The results obtained by this model are presented in Figures \ref{fig:uvbbarb}, \ref{fig:uvbhouse}, and \ref{fig:uvbleo}.

This algorithm works, but its main drawback is that it captures some structures informations (like the legs of the table in the Barbara image; see Figure \ref{fig:uvbbarb}). This behavior appears because the space $E$ is much bigger than the space $G$, in particular the space $E$ contains functions that are not only textures.

\begin{figure}[!ht]
    \centering
    %\begin{tabular}{m{5cm}m{5cm}}
    \begin{tabular}{cc}
        \includegraphics[width=0.4\textwidth]{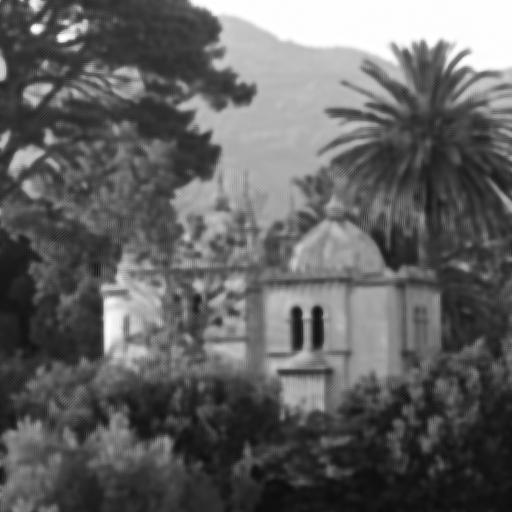} & \includegraphics[width=0.4\textwidth]{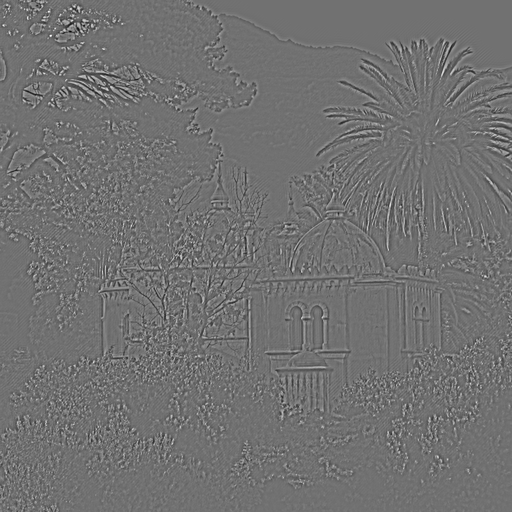} \\
        Structures                                               & Textures
    \end{tabular}
    \caption{$BV$-$H^{-1}$ structures $+$ textures image decomposition of House image.}
    \label{fig:uvshouse}
\end{figure}

\begin{figure}[!ht]
    \centering
    %\begin{tabular}{m{5cm}m{5cm}}
    \begin{tabular}{cc}
        \includegraphics[width=0.4\textwidth]{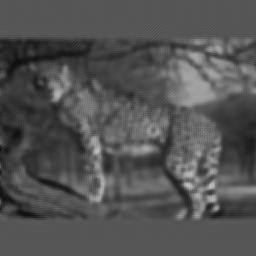} & \includegraphics[width=0.4\textwidth]{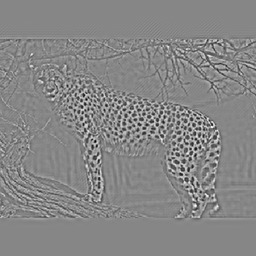} \\
        Structures                                               & Textures
    \end{tabular}
    \caption{$BV$-$H^{-1}$ structures $+$ textures image decomposition of Leopard image.}
    \label{fig:uvsleo}
\end{figure}

\cite{vese2} explore the possibility of replacing the space $G$ by the Sobolev space $H^{-1}$. They propose the following functional ($v$ is obtained by $v=f-u$):

\begin{equation}
    F_{\lambda}^{VS}(u)=J(u)+\lambda\|f-u\|_{H^{-1}}^2,
\end{equation}
where $\|v\|_{H^{-1}}=\int |\nabla (\Delta^{-1})v|^{2}dxdy$. The authors give the corresponding Euler-Lagrange equations and their discretization. Another way to numerically solve the problem is to use a modified version of Chambolle's projector $P_{H_{\lambda}^{-1}}$ (see Appendix \ref{ap:chambolle}). Figures \ref{fig:uvsbarb}, \ref{fig:uvshouse}, and \ref{fig:uvsleo} present the results obtained with this algorithm.

\begin{figure}[!ht]
    \centering
    %\begin{tabular}{m{5cm}m{5cm}}
    \begin{tabular}{cc}
        \includegraphics[width=0.4\textwidth]{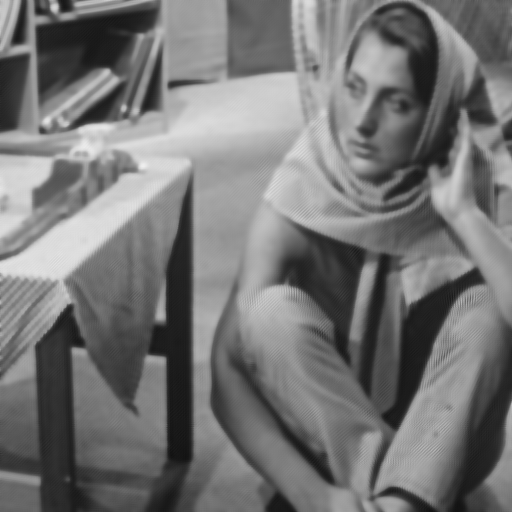} & \includegraphics[width=0.4\textwidth]{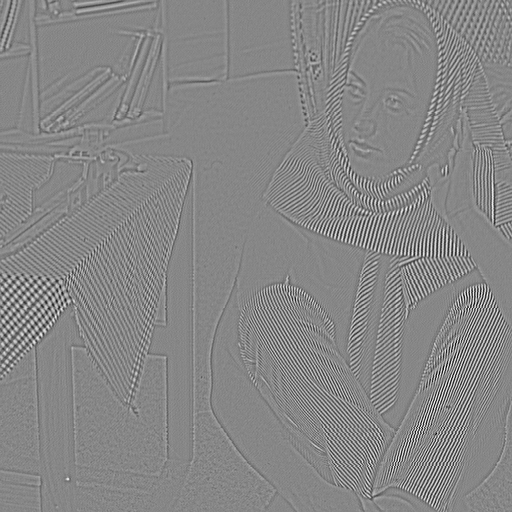} \\
        Structures                                                & Textures
    \end{tabular}
    \caption{$BV$-$H^{-1}$ structures $+$ textures image decomposition of Barbara image.}
    \label{fig:uvsbarb}
\end{figure}

Some other models were proposed that test different spaces to replace $BV$ or $G$ spaces. We mention the work of \cite{aujoluvw,aujol3} who propose replacing the space $BV$ by the smaller Besov space $B_{1,1}^1$, or replacing $G$ by some Hilbert spaces, which permits the possibility of extracting textures with a certain directionality. \cite{haddad} proposes using the Besov space $\dot{B}_{1,\infty}^1$, instead of $BV$ (the norms over these two spaces are equivalent) with the $L^2$ norm for the $v$ part. \cite{triet2,triet1} study the use of the spaces $\Div(BMO)$, $\dot{BMO}^{-\alpha}$, and $\dot{W}^{-\alpha,p}$ to modelize the textures component.
% %==============================================================================
% %  STRUCTURES + TEXTURES + BRUIT
% %==============================================================================
\section{Structures + Textures + Noise Decomposition}\label{sec:3dec}

\begin{figure}[!ht]
    \centering
    \begin{tabular}{m{5cm}m{5cm}}
        \includegraphics[width=0.4\textwidth]{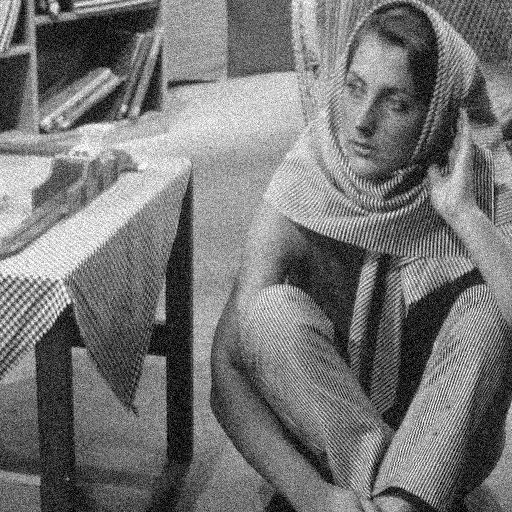} & \includegraphics[width=0.4\textwidth]{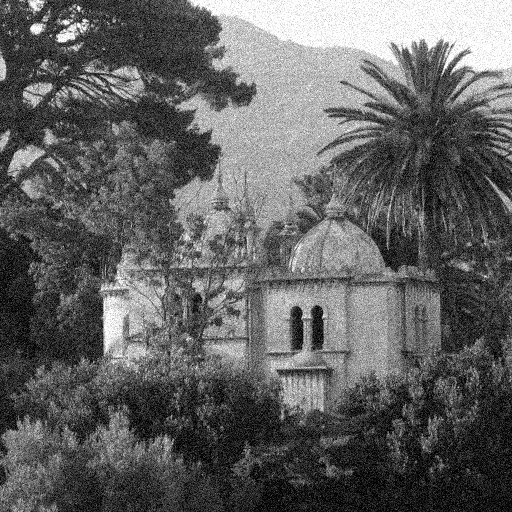} \\
        \includegraphics[width=0.4\textwidth]{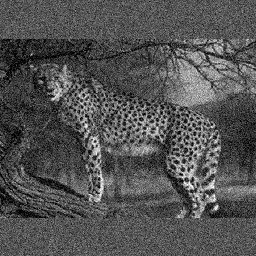} &
    \end{tabular}
    \caption{Original Barbara, House, and Leopard images corrupted by gaussian noise ($\sigma=20$).}
    \label{fig:nimorig}
\end{figure}

\begin{figure}[!ht]
    \centering
    %\begin{tabular}{m{5cm}m{5cm}}
    \begin{tabular}{cc}
        \includegraphics[width=0.4\textwidth]{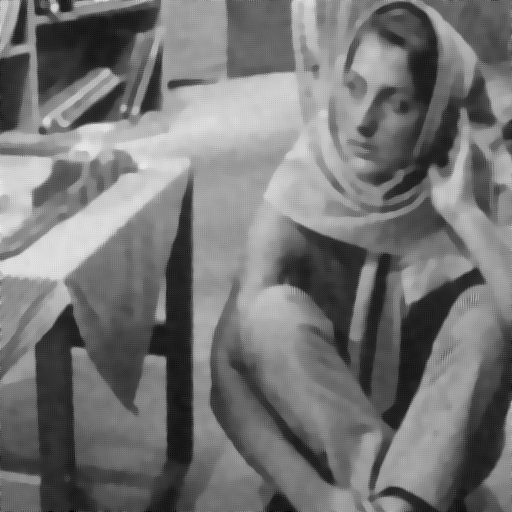} & \includegraphics[width=0.4\textwidth]{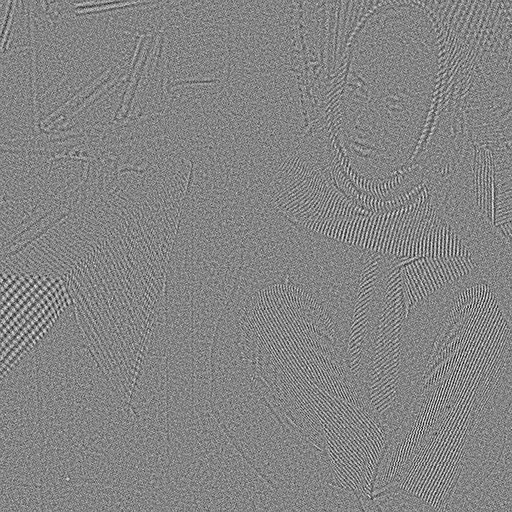} \\
        Structures                                                & Textures
    \end{tabular}
    \caption{$BV$-$G$ structures $+$ textures image decomposition of the noisy Barbara image.}
    \label{fig:uvnbarb}
\end{figure}

The previous algorithms yield good results but are of limited interest for noisy images (we add a gaussian noise with $\sigma=20$ on each test image of Figure \ref{fig:imorig}; the corresponding noisy test images can be viewed in Figure \ref{fig:nimorig}). Indeed, noise can be viewed as a very highly oscillatory function (this means that noise can be view as living in the space $G$). Therefore, the algorithms incorporate the noise in the textures components. Then the textures are corrupted by noise (see Figure \ref{fig:uvnbarb} for example).\\

In this section, we present some extension of the two-component model to the three-component model, $f=u+v+w$, which could discriminate among structures ($u$), textures ($v$), and noise ($w$).

\subsection{$BV$-$G$-$G$ Local Adaptative Model}
In \cite{jegilles2}, we proposed a new model to decompose an image into three parts: structures ($u$), textures ($v$), and noise ($w$). As in the $u+v$ model, we consider that structures and textures are modelized by functions in $BV$ and $G$ spaces, respectively. We also consider a zero mean gaussian noise added to the image. Let us view noise as a specific very oscillating function. In virtue of Meyer's work (\cite{meyer}), where it is shown that the more a function is oscillatory, the smaller its $G$-norm is, we propose to modelize $w$ as a function in $G$ and consider that its $G$-norm is much smaller than the norm of textures ($\|v\|_G  \gg \|w\|_G$). These assumptions are equivalent to choosing
\begin{equation}
    v\in G_{\mu_1}\text{ , }w\in G_{\mu_2}, \qquad \text{where} \quad \mu_1 \gg \mu_2.
\end{equation}

To increase the performance, we propose adding a local adaptability behavior to the algorithm following an idea proposed by \cite{gilboa}. These authors investigate the ROF model given by Eq.(\ref{eq:rof2bis}) and propose a modified version that can preserve textures in the denoising process. To do this, they do not choose $\lambda$ as a constant on the entire image but as a function $\lambda(f)(x,y)$ which represents local properties of the image. In a cartoon-type region, the algorithm enhances the denoising process by increasing the value of $\lambda$; in a texture-type region, the algorithm decreases $\lambda$ to attenuate the regularization to preserve the details of textures. So $\lambda(f)(x,y)$ can be viewed as a smoothed partition between textured and untextured regions.\\
Then, in order to decompose an image into three parts, we propose to use the following functional:

\begin{equation}
    F_{\lambda ,\mu_1 ,\mu_2}^{JG}(u,v,w)= J(u)+J^*\left(\frac{v}{\mu_1}\right)+J^*\left(\frac{w}{\mu_2}\right)+(2\lambda)^{-1}\|f-u-\nu_1 v-\nu_2 w\|_{L^2}^2,
\end{equation}
where the functions $\nu_i$ represent the smoothed partition of textured and untextured regions (and play the role of $\lambda$ in Gilboa's paper). The $\nu_i$ functions must have the following behavior:\\
\begin{itemize}
    \item for a textured region, we want to favor $v$ instead of $w$. This is equivalent to $\nu_1$ close to $1$ and $\nu_2$ close to $0$,
    \item for an untextured region, we want to favor $w$ instead of $v$. This is equivalent to $\nu_1$ closed to $0$ and $\nu_2$ close to $1$.\\
\end{itemize}
We see that $\nu_1$ and $\nu_2$ are complementary, so it is natural to choose $\nu_2=1-\nu_1:\mathbb{R}^2\rightarrow ]0;1[$. The choice of $\nu_1$ and $\nu_2$ is discussed after the following proposition, which characterizes the minimizers of $F_{\lambda ,\mu_1 ,\mu_2}^{JG}(u,v,w)$.

\begin{proposition}\label{prop:uvw}
    Let $u\in BV$, $v\in G_{\mu_1}$, and $w\in G_{\mu_2}$ be the structures, textures, and noise parts, respectively, and $f$ the original noisy image. Let the functions $(\nu_1(f)(.,.),\nu_2(f)(.,.))$ be defined on $\mathbb{R}^2\rightarrow ]0;1[$, and assume that these functions could be considered as locally constant compared to the variation of $v$ and $w$. Then a minimizer defined by
    \begin{equation}\label{equ:uvw2}
        (\hat{u},\hat{v},\hat{w})=\underset{(u,v,w)\in BV\times G_{\mu_1} \times G_{\mu_2}}{\arg}\min F_{\lambda ,\mu_1 ,\mu_2}^{JG}(u,v,w),
    \end{equation}
    is given by
    \begin{align}
        \hat{u} & =f-\nu_1\hat{v}-\nu_2\hat{w}-P_{G_{\lambda}}(f-\nu_1\hat{v}-\nu_2\hat{w}), \\
        \hat{v} & =P_{G_{\mu_1}}\left(\frac{f-\hat{u}-\nu_2\hat{w}}{\nu_1}\right),           \\
        \hat{w} & =P_{G_{\mu_2}}\left(\frac{f-\hat{u}-\nu_1\hat{v}}{\nu_2}\right),
    \end{align}
    where $P_{G_{\mu}}$ denotes Chambolle's non-linear projectors (see Appendix \ref{ap:chambolle}).\\
\end{proposition}

The proof of this proposition can be found in \cite{jegilles2}. As in the two-part $BV$-$G$ decomposition model, we get an equivalent numerical scheme:\\

\begin{center}
    %\fbox{\parbox[h]{0.9\textwidth}{
    \begin{enumerate}
        \item Initialization: $u_0=v_0=w_0=0$,
        \item Compute $\nu_1$ and $\nu_2=1-\nu_1$ from $f$,
        \item Compute $w_{n+1}=P_{G_{\mu_2}}\left(\frac{f-u_n-\nu_1v_n}{\nu_2+\kappa}\right)$, ($\kappa$ is a small value in order to prevent the division by zero),
        \item Compute $v_{n+1}=P_{G_{\mu_1}}\left(\frac{f-u_n-\nu_2w_{n+1}}{\nu_1+\kappa}\right)$,
        \item Compute $u_{n+1}=f-\nu_1v_{n+1}-\nu_2w_{n+1}-P_{G_{\lambda}}(f-\nu_1v_{n+1}-\nu_2w_{n+1})$,
        \item If $\max\{|u_{n+1}-u_n|,|v_{n+1}-v_n|,|w_{n+1}-w_n|\}\leqslant\epsilon$ or if we did $N_{step}$ \\ iterations then stop the algorithm, else jump to step 3.
    \end{enumerate}%}}
\end{center}

\begin{figure}[t!]
    \begin{center}
        \includegraphics[width=0.3\textwidth]{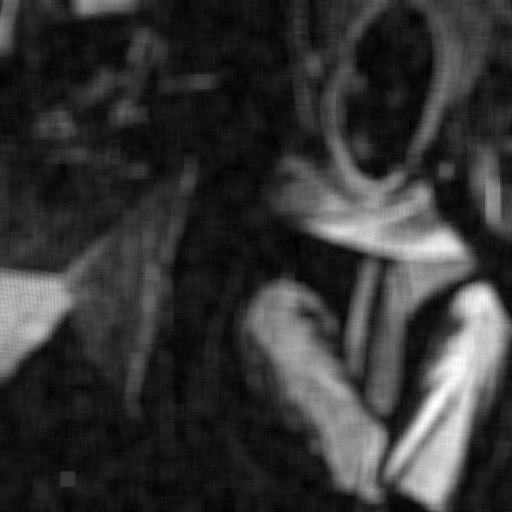}
        \caption{Texture partition $\nu_1$ obtained by local variance computation.}
        \label{fig:nu}
    \end{center}
\end{figure}

Concerning the choice of the $\nu_i$ functions, we were inspired by the work of \cite{gilboa}. The authors choose to compute a local variance on the texture + noise part of the image obtained by the ROF model ($f-u$). In our model, we use the same strategy but on the $v$ component obtained by the two parts decomposition algorithm. This choice is implied by the fact that the additive gaussian noise can be considered as orthogonal to textures. As a consequence, the variance of a textured region is larger than the variance of an untextured region.\\
So, in practice, we first compute the two-part decomposition of the image $f$. On the textures part, for all the pixels $(i,j)$, we compute the local variance on a small window (odd size $L$) centered on $(i,j)$. At the least, we normalized it to obtain the values in $]0;1[$. All the details about the computation of the $\nu_i$'s function can be found in \cite{jegilles2}. Figure \ref{fig:nu} shows an example from the noisy Barbara image. As expected, the variance is higher in the textured regions and lower in the others.

\begin{figure}[!ht]
    \centering
    %\begin{tabular}{m{5cm}m{5cm}}
    \begin{tabular}{cc}
        \includegraphics[width=0.4\textwidth]{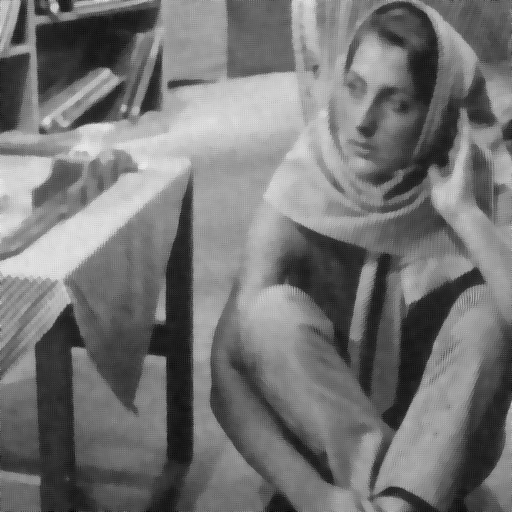} & \includegraphics[width=0.4\textwidth]{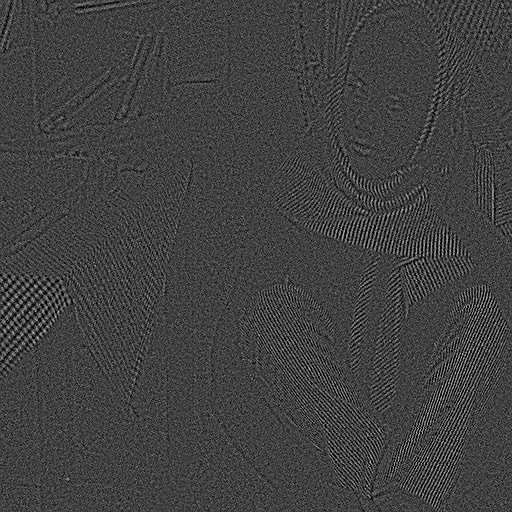} \\
        Structures                                                         & Textures                                                           \\
        \includegraphics[width=0.4\textwidth]{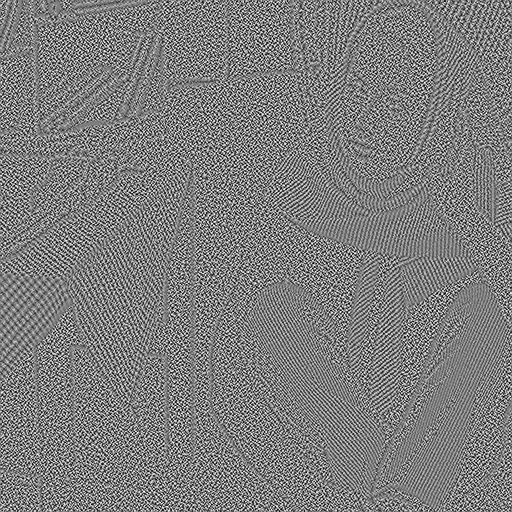} &                                                                    \\
        Noise
    \end{tabular}
    \caption{$BV$-$G$-$G$ structures $+$ textures $+$ noise image decomposition of Barbara image.}
    \label{fig:uvwjgbarb}
\end{figure}

\begin{figure}[!ht]
    \centering
    %\begin{tabular}{m{5cm}m{5cm}}
    \begin{tabular}{cc}
        \includegraphics[width=0.4\textwidth]{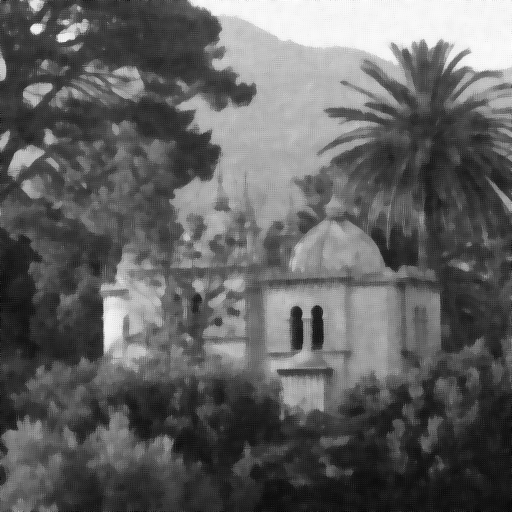} & \includegraphics[width=0.4\textwidth]{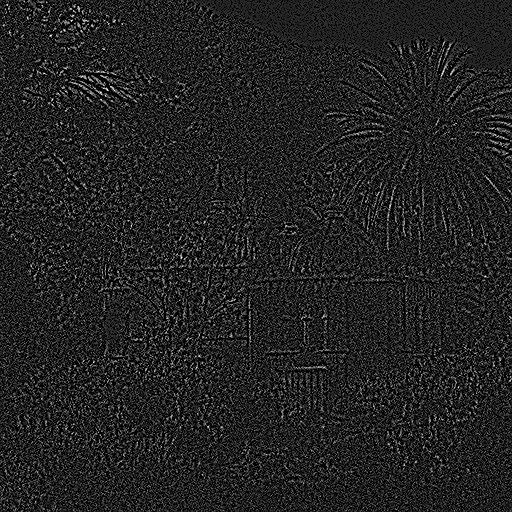} \\
        Structures                                                        & Textures                                                          \\
        \includegraphics[width=0.4\textwidth]{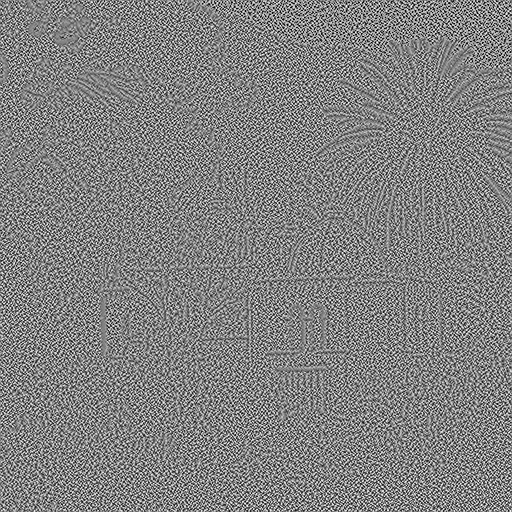} &                                                                   \\
        Noise
    \end{tabular}
    \caption{$BV$-$G$-$G$ structures $+$ textures $+$ noise image decomposition of House image.}
    \label{fig:uvwghouse}
\end{figure}

\begin{figure}[!ht]
    \centering
    %\begin{tabular}{m{5cm}m{5cm}}
    \begin{tabular}{cc}
        \includegraphics[width=0.4\textwidth]{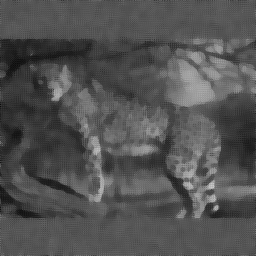} & \includegraphics[width=0.4\textwidth]{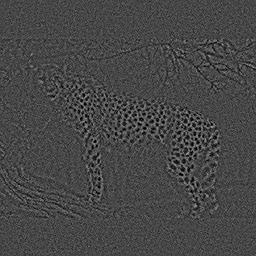} \\
        Structures                                                        & Textures                                                          \\
        \includegraphics[width=0.4\textwidth]{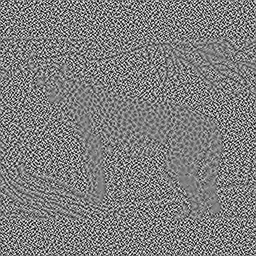} &                                                                   \\
        Noise
    \end{tabular}
    \caption{$BV$-$G$-$G$ structures $+$ textures $+$ noise image decomposition of Leopard image.}
    \label{fig:uvwjgleo}
\end{figure}

Figures \ref{fig:uvwjgbarb}, \ref{fig:uvwghouse}, and \ref{fig:uvwjgleo} show the results of the $u+v+w$ decomposition we obtained by the $BV$-$G$-$G$ local adaptive model. This model can separate noise from the textures. If we look more precisely, we can see that some residual noise remains in the textures, and some textures are partially captured in the noise part. This is due to the choice of the parameters $\lambda,\mu_1$, and $\mu_2$ which act on the separability power of the algorithm.

\subsection{Aujol-Chambolle $BV$-$G$-$E$ Model}
The same time as our work, \cite{aujoluvw} thought of the same structures + textures + noise decomposition problem. They proposed a model close to our model described in the previous subsection but with the difference that they consider the noise as a distribution taken in the Besov space $E=\dot{B}_{-1,\infty}^{\infty}$. Then the associated functional is

\begin{equation}\label{eqn:aujoluvw}
    F_{\lambda,\mu,\delta}^{AC2}(u,v,w)= J(u)+J^*\left(\frac{v}{\mu}\right)+B^*\left(\frac{w}{\delta}\right)+(2\lambda)^{-1}\|f-u-v-w\|_{L^2}^2,
\end{equation}
where $u\in BV$, $v\in G_{\mu}$, and $w\in E_{\delta}$ as defined in the previous sections. The authors prove that the minimizers are (see \cite{aujoluvw}):
\begin{align}
     & \hat{u}=f-\hat{v}-\hat{w}-P_{G_{\lambda}}(f-\hat{v}-\hat{w}),                               \\
     & \hat{v}=P_{G_{\mu}}(f-\hat{u}-\hat{w}),                                                     \\
     & \hat{w}=P_{E_{\delta}}(f-\hat{u}-\hat{v})=f-\hat{u}-\hat{v}-WST(f-\hat{u}-\hat{v},2\delta),
\end{align}
where $WST(f-\hat{u}-\hat{v},2\delta)$ is the Wavelet Soft Thresholding operator applied on $f-\hat{u}-\hat{v}$ with a threshold set to $2\delta$.

Then the numerical algorithm is given by

\begin{center}
    %\fbox{\parbox[h]{0.9\textwidth}{
    \begin{enumerate}
        \item Initialization: $u_0=v_0=w_0=0$,
        \item Compute $w_{n+1}=f-u_n-v_n-WST(f-u_n-v_n,2\delta)$,
        \item Compute $v_{n+1}=P_{G_{\mu}}(f-u_n-w_{n+1})$,
        \item Compute $u_{n+1}=f-v_{n+1}-w_{n+1}-P_{G_{\lambda}}(f-v_{n+1}-w_{n+1})$,
        \item If $\max\{|u_{n+1}-u_n|,|v_{n+1}-v_n|,|w_{n+1}-w_n|\}\leqslant\epsilon$ or if we performed $N_{step}$ \\ iterations, then stop the algorithm, else jump to step 2.
    \end{enumerate}%}}
\end{center}

\begin{figure}[!ht]
    \centering
    %\begin{tabular}{m{5cm}m{5cm}}
    \begin{tabular}{cc}
        \includegraphics[width=0.4\textwidth]{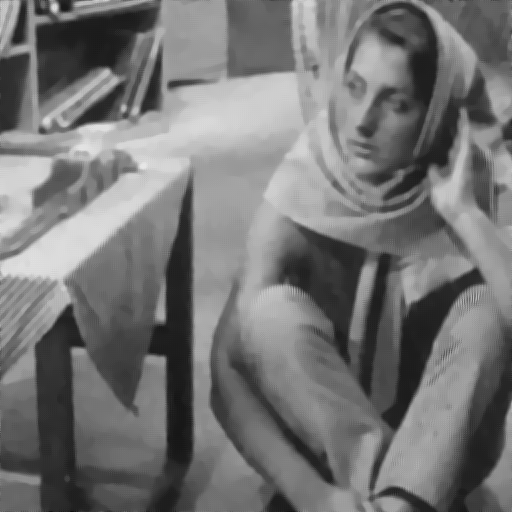} & \includegraphics[width=0.4\textwidth]{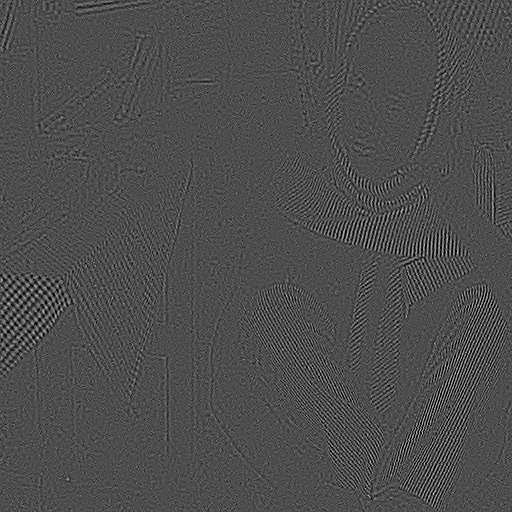} \\
        Structures                                                            & Textures                                                              \\
        \includegraphics[width=0.4\textwidth]{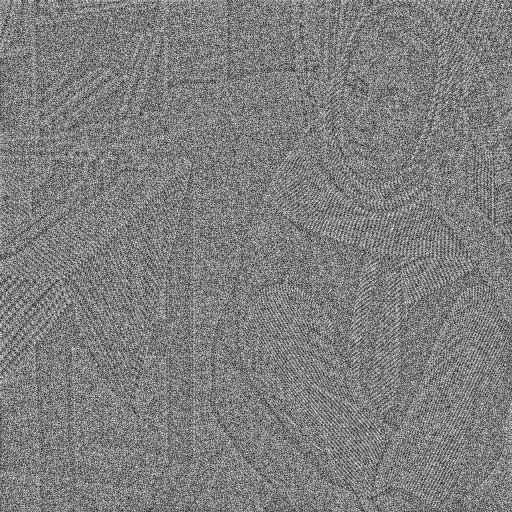} &                                                                       \\
        Noise
    \end{tabular}
    \caption{$BV$-$G$-$E$ structures $+$ textures $+$ noise image decomposition of Barbara image.}
    \label{fig:uvwaubarb}
\end{figure}

\begin{figure}[!ht]
    \centering
    %\begin{tabular}{m{5cm}m{5cm}}
    \begin{tabular}{cc}
        \includegraphics[width=0.4\textwidth]{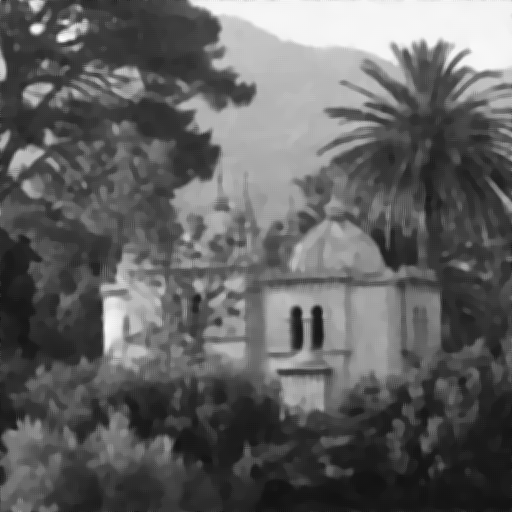} & \includegraphics[width=0.4\textwidth]{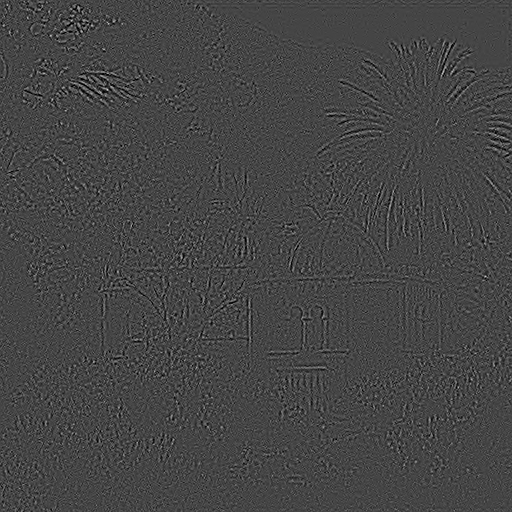} \\
        Structures                                                           & Textures                                                             \\
        \includegraphics[width=0.4\textwidth]{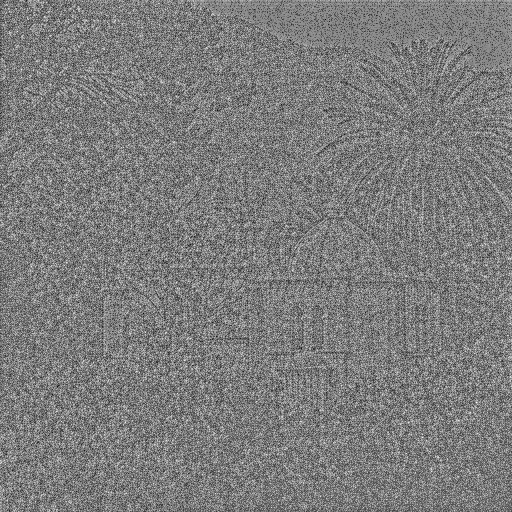} &                                                                      \\
        Noise
    \end{tabular}
    \caption{$BV$-$G$-$E$ structures $+$ textures $+$ noise image decomposition of House image.}
    \label{fig:uvwauhouse}
\end{figure}

\begin{figure}[!ht]
    \centering
    %\begin{tabular}{m{5cm}m{5cm}}
    \begin{tabular}{cc}
        \includegraphics[width=0.4\textwidth]{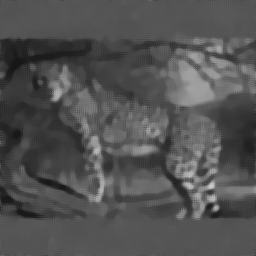} & \includegraphics[width=0.4\textwidth]{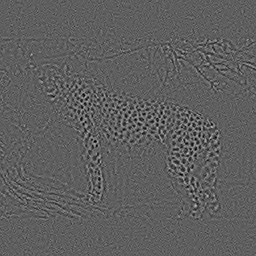} \\
        Structures                                                           & Textures                                                             \\
        \includegraphics[width=0.4\textwidth]{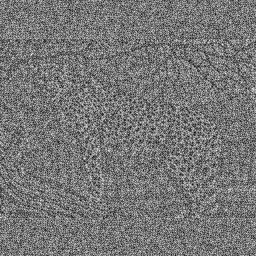} &                                                                      \\
        Noise
    \end{tabular}
    \caption{$BV$-$G$-$E$ structures $+$ textures $+$ noise image decomposition of Leopard image.}
    \label{fig:uvwauleo}
\end{figure}

The results of this algorithm on our test images are shown in Figures \ref{fig:uvwaubarb}, \ref{fig:uvwauhouse}, and \ref{fig:uvwauleo}, respectively. We can see that textures are better denoised by this model. This is a consequence of a better noise modeling by distributions in the Besov space. But the residual texture is more important than the one given by our algorithm in the noise part. Another drawback appears in the structures part; the edges in the image are damaged because some important wavelet coefficients are removed. Previously, \cite{jegilles2} provides the possibility to add the local adaptivity behavior of the $BV$-$G$-$G$ model to the $BV$-$G$-$E$ model. We refer the reader to \cite{jegilles2} to see the $BV$-$G$-$E$ local adaptivity functional and find the corresponding results. This modified version shows less improvement compared to the original. We prefer to explore the replacement of wavelets by new geometric multiresolution tools such as contourlets.

\subsection{The $BV$-$G$-$\dot{Co}_{-1,\infty}^{\infty}$ Decomposition Model}\label{sec:contdec}
As mentionned previously, the new directional multiresolution tools, such as curvelets or contourlets, exhibit very good results in denoising. They also better reconstruct the edge in an image. So, the idea to replace the use of wavelet by curvelets or contourlets naturally arises. In this paper, we focus on the choice of contourlets. This choice is equivalent to changing the Besov space in the model described in the previous subsection by the homogeneous contourlet space $\dot{Co}_{-1,\infty}^{\infty}$. Then, the equivalent functional is given in Eq.(\ref{eqn:uvwct}) as below:

\begin{equation}\label{eqn:uvwct}
    F_{\lambda,\mu,\delta}^{Co}(u,v,w)=J(u)+J^*\left(\frac{v}{\mu}\right)+J_{Co}^*\left(\frac{w}{\delta}\right)+(2\lambda)^{-1}\|f-u-v-w\|_{L^2}^2,
\end{equation}
where $J_{Co}^*(f)$ is the indicator function over the set $Co_1$ if we denote $Co_{\delta}=\left\{f\in Co_{-1,\infty}^{\infty} / \|f\|_{Co_{-1,\infty}^{\infty}}\leqslant \delta \right\}$ (norm over the contourlet spaces is defined in the subsection \ref{sec:contspace}) defined by
\begin{equation}
    J_{Co}^*(f)=
    \begin{cases}
        0 \qquad \text{if} \; f\in Co_1 \\
        +\infty \qquad \text{else.}
    \end{cases}
\end{equation}

Then, the following proposition gives the solutions that minimize the previous functional.
\begin{proposition}\label{pro:uvwjg3}
    Let $u\in BV$, $v\in G_{\mu}$, $w\in Co_{\delta}$ be the structures, textures, and noise components derived from the image decomposition. Then the solution
    \begin{equation}
        (\hat{u},\hat{v},\hat{w})=\underset{(u,v,w)\in BV\times G_{\mu}\times Co_{\delta}}{\arg} \inf F_{\lambda,\mu,\delta}^{Co}(u,v,w)
    \end{equation}
    is given by
    \begin{align*}
        \hat{u} & =f-\hat{v}-\hat{w}-P_{G_{\lambda}}(f-\hat{v}-\hat{w})         \\
        \hat{v} & =P_{G_{\mu}}\left(f-\hat{u}-\hat{w}\right)                    \\
        \hat{w} & =f-\hat{u}-\hat{v}-CST\left(f-\hat{u}-\hat{v};2\delta\right),
    \end{align*}
    where $P_{G_{\lambda}}$ is the Chambolle nonlinear projector and $CST(f,2\delta)$ is the contourlet soft thresholding operator of $f-u-v$.\\
\end{proposition}

\begin{proof}
    The components $\hat{u}, \hat{v}$ are obtained by the same arguments used in the proof of Proposition \ref{prop:uvw} (this proof is available in \cite{jegilles2}). The particular point concerns the expression of $\hat{w}$ expressed with the soft thresholding of the contourlet coefficients. Assume we want to minimize $F_{\lambda,\mu,\delta}^{Co}(u,v,w)$ compared to $w$; it is equivalent to find $w$ solution of (we set $g=f-u-v$)
    \begin{equation}
        \hat{w}=\underset{w\in Co_{\delta}}{\arg}\min \left\lbrace \|g-w\|_{L^2}^2\right\rbrace.
    \end{equation}
    We can replace it by its dual formulation: $\hat{w}=g-\hat{h}$, such that
    \begin{equation}\label{eq:dualct}
        \hat{h}=\underset{h\in Co_{1,1}^1}{\arg}\min \left\lbrace 2\delta\|h\|_{Co_{1,1}^1}+\|g-h\|_{L^2}^2\right\rbrace.
    \end{equation}
    We can use the same approach used by \cite{chambolle2}.

    Let $(c_{j,k,n})_{j\in\Z , 0\leqslant k\leqslant 2^{(l_j)},n\in\Z^2}$ and $(d_{j,k,n})_{j\in\Z , 0\leqslant k\leqslant 2^{(l_j)},n\in\Z^2}$ denote the coefficients issued from the contourlet expansions of $g$ and $h$, respectively. As contourlets form a tight frame, with a bound of $1$, we have (we denote $\Omega =\Z \times \llbracket 0,2^{(l_j)} \rrbracket\times \Z^2$)
    \begin{equation}
        \|g\|_{L_2}^2=\sum_{(j,k,n)\in\Omega}|c_{j,k,n}|^2.
    \end{equation}

    Then Eq.(\ref{eq:dualct}) can be rewritten as
    \begin{equation}
        \sum_{(j,k,n)\in \Omega}|c_{j,k,n}-d_{j,k,n}|^2+2\delta\sum_{(j,k,n)\in \Omega}|d_{j,k,n}|,
    \end{equation}
    which is equivalent to
    \begin{equation}
        |c_{j,k,n}-d_{j,k,n}|^2+2\delta|d_{j,k,n}|.
    \end{equation}
    However, \cite{chambolle2} prove that the solution of this kind of problem is the soft thresholding of the coefficients $(c_{j,k,n})$ with $2\delta$ as the threshold.

    Then $\hat{h}=CST(g,2\delta)$, which by duality implies that $\hat{w}=g-CST(g,2\delta)$. We conclude that
    \begin{equation}
        \hat{w}=f-\hat{u}-\hat{v}-CST(f-\hat{u}-\hat{v},2\delta)
    \end{equation}
    which end the proof.
\end{proof}

The corresponding numerical scheme is the same as in the $BV$-$G$-$E$ algorithm, except we replace the wavelet expansion by the contourlet expansion in the soft thresholding:\\

\begin{center}
    %\fbox{\parbox[h]{0.9\textwidth}{
    \begin{enumerate}
        \item Initialization: $u_0=v_0=w_0=0$,
        \item Compute $w_{n+1}=f-u_n-v_n-CST(f-u_n-v_n,2\delta)$,
        \item Compute $v_{n+1}=P_{G_{\mu}}(f-u_n-w_{n+1})$,
        \item Compute $u_{n+1}=f-v_{n+1}-w_{n+1}-P_{G_{\lambda}}(f-v_{n+1}-w_{n+1})$,
        \item If $\max\{|u_{n+1}-u_n|,|v_{n+1}-v_n|,|w_{n+1}-w_n|\}\leqslant\epsilon$ or if we performed $N_{step}$ \\ iterations, then stop the algorithm; else jump to step 2.
    \end{enumerate}%}}
\end{center}

\begin{figure}[!ht]
    \centering
    %\begin{tabular}{m{5cm}m{5cm}}
    \begin{tabular}{cc}
        \includegraphics[width=0.4\textwidth]{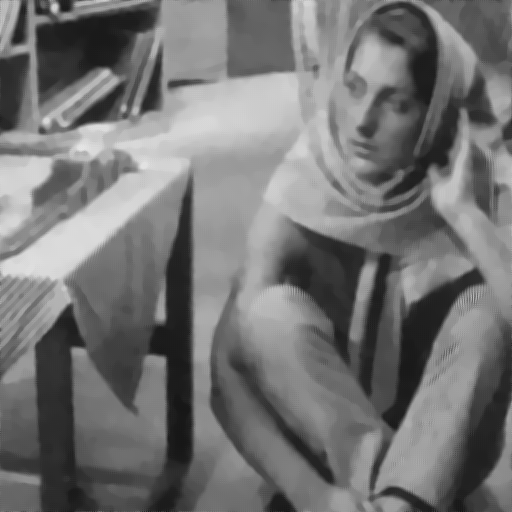} & \includegraphics[width=0.4\textwidth]{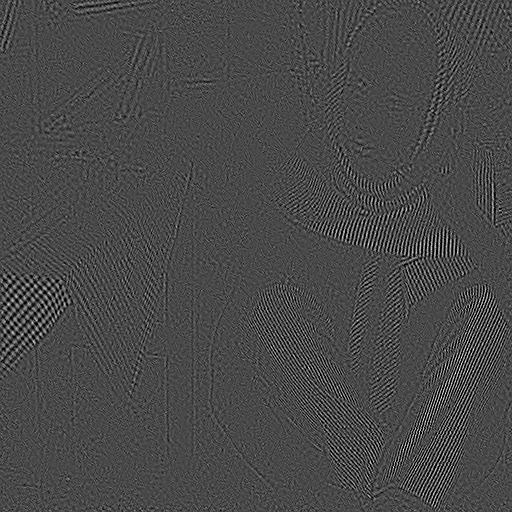} \\
        Structures                                                        & Textures                                                          \\
        \includegraphics[width=0.4\textwidth]{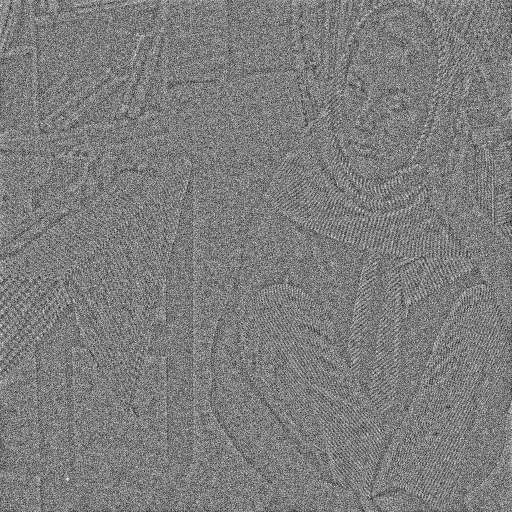} &                                                                   \\
        Noise
    \end{tabular}
    \caption{$BV$-$G$-$Co$ structures $+$ textures $+$ noise image decomposition of Barbara image.}
    \label{fig:uvwcobarb}
\end{figure}

\begin{figure}[!ht]
    \centering
    %\begin{tabular}{m{5cm}m{5cm}}
    \begin{tabular}{cc}
        \includegraphics[width=0.4\textwidth]{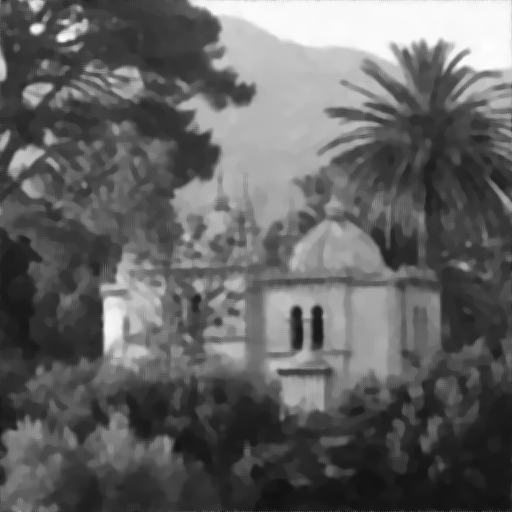} & \includegraphics[width=0.4\textwidth]{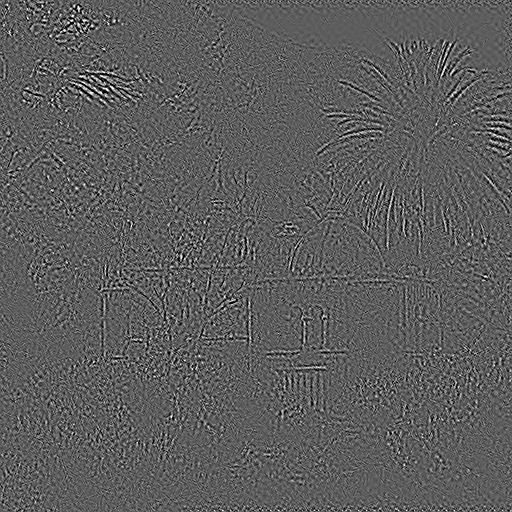} \\
        Structures                                                       & Textures                                                         \\
        \includegraphics[width=0.4\textwidth]{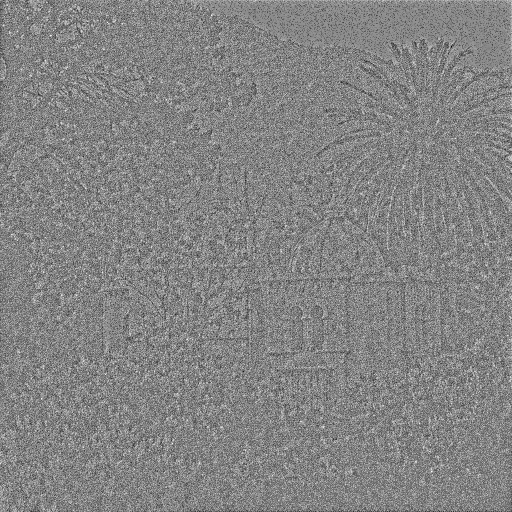} &                                                                  \\
        Noise
    \end{tabular}
    \caption{$BV$-$G$-$Co$ structures $+$ textures $+$ noise image decomposition of House image.}
    \label{fig:uvwcobat}
\end{figure}

\begin{figure}[!ht]
    \centering
    %\begin{tabular}{m{5cm}m{5cm}}
    \begin{tabular}{cc}
        \includegraphics[width=0.4\textwidth]{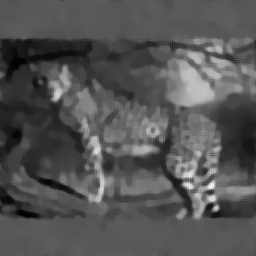} & \includegraphics[width=0.4\textwidth]{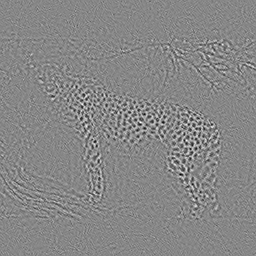} \\
        Structures                                                       & Textures                                                         \\
        \includegraphics[width=0.4\textwidth]{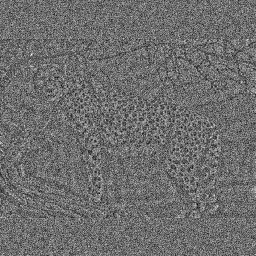} &                                                                  \\
        Noise
    \end{tabular}
    \caption{$BV$-$G$-$Co$ structures $+$ textures $+$ noise image decomposition of Leopard image.}
    \label{fig:uvwcoleo}
\end{figure}

\begin{figure}[!ht]
    \centering
    %\begin{tabular}{m{5cm}m{5cm}}
    \begin{tabular}{cc}
        \includegraphics[width=0.4\textwidth]{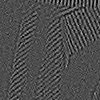} & \includegraphics[width=0.4\textwidth]{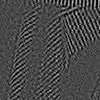} \\
        Wavelet thresholding                              & Contourlet thresholding
    \end{tabular}
    \caption{Zoomed image for the textured components of wavelet, and contourlet, based algorithms.}
    \label{fig:zoomuvw}
\end{figure}

Figures \ref{fig:uvwcobarb}, \ref{fig:uvwcobat}, and \ref{fig:uvwcoleo} show the results obtained by replacing wavelets by contourlets. The advantage of using geometric frames is that it preserves well the integrity of oriented textures as seen in the zoomed images in Figure \ref{fig:zoomuvw}.\\

In this section, we presented many decomposition models. We can imagine the use of other frames and basis like curvelets, cosines, and so on. The idea of decomposing an image by thresholding different basis expansion coefficients corresponds to the recent theory of morphological component analysis (MCA) \cite{starck2}, \cite{starck1}. This approach seeks sparse representation of the different components and is useful for sources separation.

% %==============================================================================
% %  EVALUATION
% %==============================================================================
\section{Performance Evaluation}\label{sec:eval}

The previous section described different decomposition models based on specific function spaces. But one question arises: Which is the best one?\\

This section adresses this question by defining well-adapted criteria and their associated metrics. We build a special test image by creating different components separatly and then by adding them. We will denote $f_0$ the test image composed of $u_0$ (the structures reference image) + $v_0$ (the textures reference image) + $w_0$ (the noise reference image). We finish by giving the measures obtained for this image.

\subsection{Test Image}\label{subsec:testim}
\begin{figure}[!ht]
    \begin{center}
        \begin{tabular}{ccc}
            \includegraphics[width=0.3\textwidth]{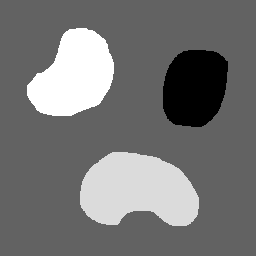} & \includegraphics[width=0.3\textwidth]{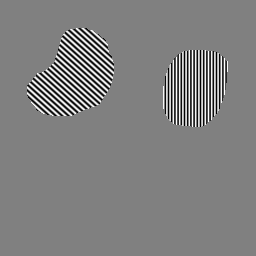} & \includegraphics[width=0.3\textwidth]{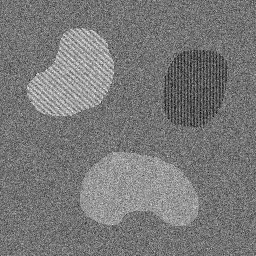} \\
        \end{tabular}
    \end{center}
    \caption{Structures and textures reference images and the recomposed test image.}
    \label{fig:refim}
\end{figure}

Because we want to compare the quality of each extracted components, we will create specific components: $u_0$ for structures, $v_0$ for textures, and $w_0$ for noise. Textures are built by sine functions over some finite domains; structures are made by drawing some shapes with an adapted software like GIMP. The noise part is simply a gaussian noise with $\sigma=20$. The $u_0$ and $v_0$ reference parts and the recomposed test image are shown in Figure \ref{fig:refim}.

\subsection{Evaluation Metrics}\label{subsec:metric}
Assume the test image is composed of known reference images $u_0$, $v_0$, and $w_0$. We choose the following criteria to measure the decomposition quality: the $L^2$-norms of errors $u-u_0$ and $v-v_0$, where $u$ and $v$ are the structures and textures components issued from the decomposition. Another quantity that is interesting to evaluate is the residual structures + textures present in the noise component $w$. To measure this quantity we prove the following proposition.

\begin{proposition}
    Let $b(i,j)$ denote a gaussian noise of variance $\sigma^2$ and $d(i,j)$ an image free of noise (we assume that the intercorrelation between $b$ and $d$ is negligible). Let $f=Ad+b$ be a simulated noise + residue image where $A\in \R$ corresponds to residue level. Then
    \begin{equation}
        \|\gamma_f-\gamma_b\|_{L^2} \approx A^2,
    \end{equation}
    where $\gamma_f$ and $\gamma_b$ are the autocorrelation functions of $f$ and $b$, respectively.
\end{proposition}

\begin{proof}
    We start by calculating the autocorrelation function of $f$:

    \begin{equation}
        \gamma_f(k,l)=\sum_{(i,j)\in \Z^2}f(i,j)f^*(i+k,j+l).
    \end{equation}
    However, we assume that images are real, then $f(i,j)=f^*(i,j)$ and we deduce that
    \begin{align}
        \gamma_f(k,l) & =\sum_{(i,j)\in \Z^2}\left[Ad(i,j)+b(i,j)\right]\left[Ad(i+k,j+l)+b(i+k,j+l)\right]   \\
                      & =\sum_{(i,j)\in \Z^2}A^2d(i,j)d(i+k,j+l)+\sum_{(i,j)\in \Z^2}b(i,j)b(i+k,j+k)+ \notag \\
                      & \sum_{(i,j)\in \Z^2}\left[Ad(i,j)b(i+k,j+l)+Ad(i+k,j+l)b(i,j)\right]                  \\
                      & =A^2\gamma_d(k,l)+\gamma_b(k,l)+A\left(\gamma_{db}(k,l)+\gamma_{bd}(k,l)\right)
    \end{align}

    Now we examine the norm $\|.\|_{L^2}$ of this autocorrelation function. First, notice that $\gamma_b(k,l)=\sigma^2\delta(k,l)$ (where $\delta(k,l)$ is the Kronecker symbol) because we assumed that the noise is gaussian. The statement of the proposition assumed that the intercorrelations are negligible; in pratice, it is easy to check that the quantity $A\left(\gamma_{db}(k,l)+\gamma_{bd}(k,l)\right)$ is negligible compared to $A^2\gamma_d(k,l)$. We deduce that
    \begin{equation}
        \gamma_f(k,l) - \gamma_b(k,l) \approx A^2\gamma_d(k,l);
    \end{equation}
    then, by passing to the norm, we get

    \begin{equation}
        \|\gamma_f-\gamma_b\|_{L^2} \approx A^2\|\gamma_d\|_{L^2}.
    \end{equation}
\end{proof}

\begin{figure}[!ht]
    \begin{center}
        \includegraphics[width=0.3\textwidth]{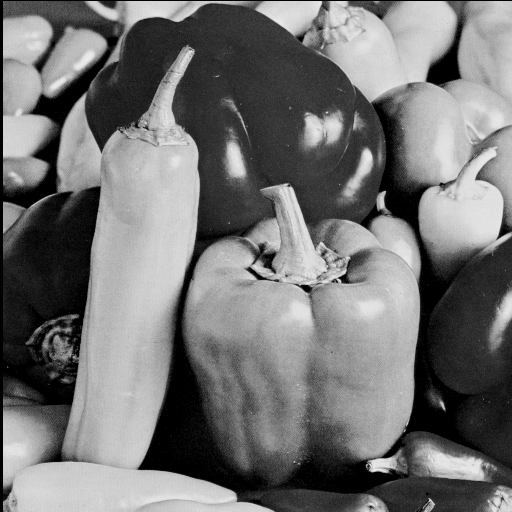}
    \end{center}
    \caption{Residual reference image.}
    \label{fig:testdenoising}
\end{figure}

\begin{figure}[!ht]
    \centering
    \begin{tabular}{ccc}
        \includegraphics[scale=0.22]{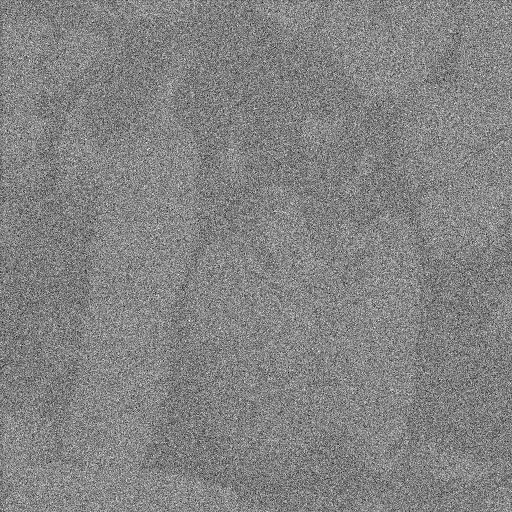} & \includegraphics[scale=0.22]{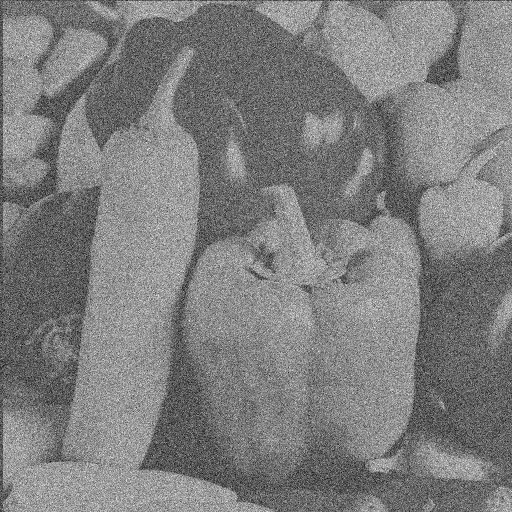} & \includegraphics[scale=0.22]{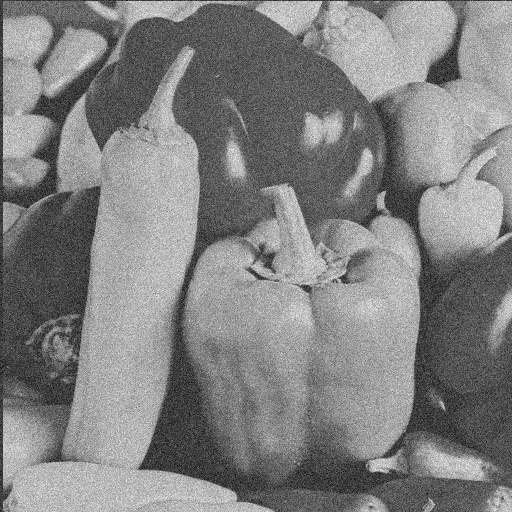} \\
        \includegraphics[scale=0.2]{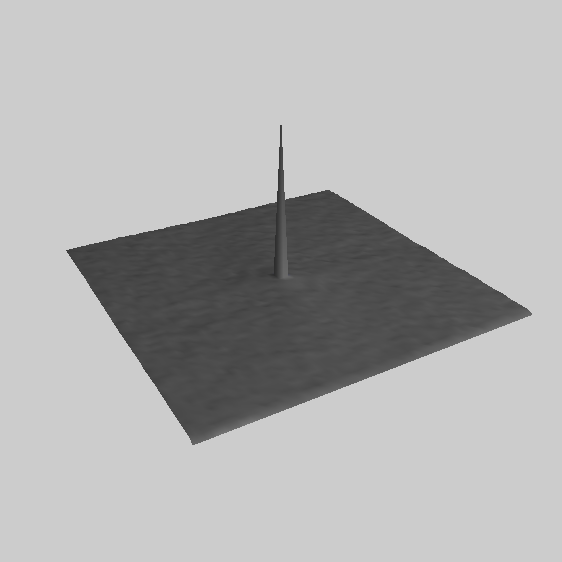}    & \includegraphics[scale=0.2]{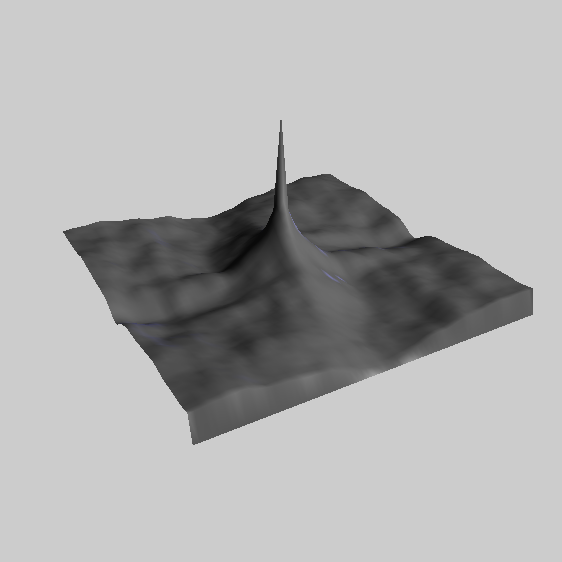}     & \includegraphics[scale=0.2]{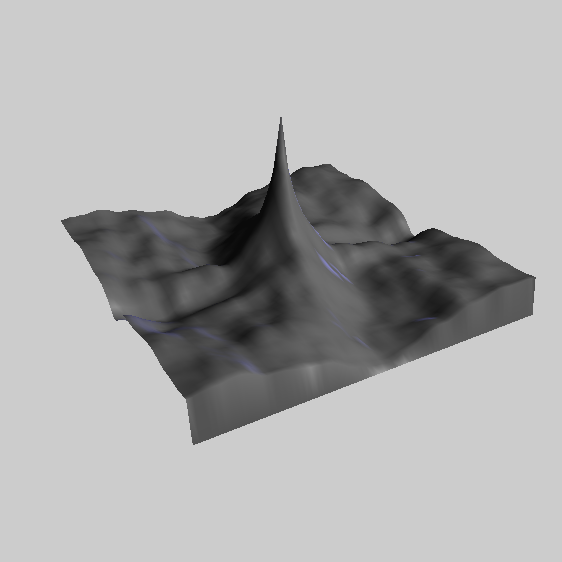}     \\
        $A=0.05$                              & $A=0.3$                               & $A=0.8$
    \end{tabular}
    \caption{Noisy reference images affected by different residual levels and their associated autocorrelation images.}
    \label{fig:testresidu}
\end{figure}

To illustrate this proposition, assume that we take the image in Figure \ref{fig:testdenoising} as $d(i,j)$ and we generate an image $b(i,j)$ full of gaussian noise ($\sigma=20$). Then we compose the image $f=Ad+b$ for the different values $A\in\{0.05;0.1;0.2;0.3;0.4;$ $0.5;0.6;0.7;0.8;$ $0.9\}$ (this means that more and more residue appears as $A$ increases, see Figure \ref{fig:testresidu} top row).

\begin{figure}[!ht]
    \begin{center}
        \begin{tabular}{m{4cm}m{8cm}}

            \begin{tabular}{|c|c|} \hline
                $A$  & $\|\gamma_f-\gamma_b\|_{L^2}$ \\ \hline
                0.05 & 849.093432                    \\ \hline
                0.1  & 3312.071022                   \\ \hline
                0.2  & 13099.095280                  \\ \hline
                0.3  & 29367.800483                  \\ \hline
                0.4  & 52118.223554                  \\ \hline
                0.5  & 81350.371724                  \\ \hline
                0.6  & 117064.247377                 \\ \hline
                0.7  & 159259.851531                 \\ \hline
                0.8  & 207937.184693                 \\ \hline
                0.9  & 263096.247142                 \\ \hline
            \end{tabular} &
            \centering\includegraphics[width=7cm]{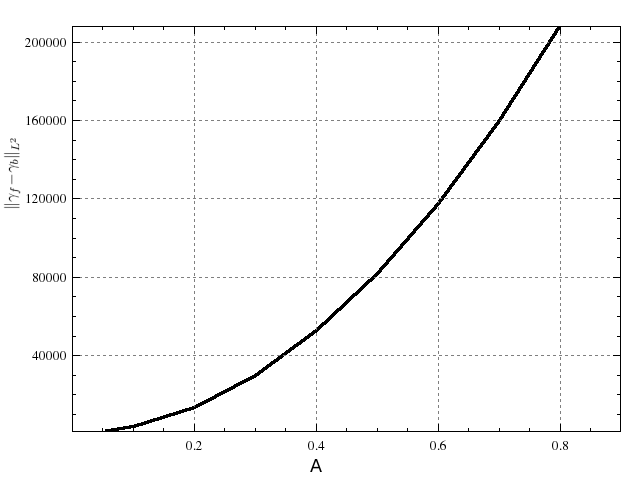}
        \end{tabular}
    \end{center}
    \caption{Results of the measure norm $\|\gamma_f-\gamma_b\|_{L^2}$ for the different values of $A$ (left) and its associated graph.}
    \label{fig:normresidu}
\end{figure}

Figure \ref{fig:normresidu} gives the measured values of $\|\gamma_f-\gamma_b\|_{L^2}$ and shows the associated graph. As announced by the proposition, we show the quadratic behavior of the norm of the autocorrelation differences as $A$ grows. We will use this metric in the next subsection to evaluate the residual quantity in the noise parts at the output of the different decomposition algorithms.

\subsection{Image Decomposition Performance Evaluation}
In this subsection we apply three-part image decomposition on the test image built in subsection \ref{subsec:testim} and use the metrics defined in subsection \ref{subsec:metric} to evaluate their performances. In this chapter, we restrict the choice of the different parameters to only the ones that give the best visual performances, but in the future, a more global, in terms of parameters variability, test could be to explore the complete behaviors of the algorithms. The choosen parameters are

\begin{flushleft}
    \begin{itemize}
        \item Algorithm $F^{JG}$: $\lambda=10$, $\mu_1=1000$, $\mu_2=100$, and a window size of $3\times 3$ pixels,
        \item Algorithm $F^{AC2}$: $\lambda=1$, $\mu=500$ and $\delta=9.4$ ($\kappa=0.2$ and $\sigma=20$),
        \item Algorithm $F^{Co}$: $\lambda=1$, $\mu=500$ and $\delta=23.5$ ($\kappa=0.5$ and $\sigma=20$).\\
    \end{itemize}
\end{flushleft}

\begin{figure}[!ht]
    \begin{center}
        \begin{tabular}{ccc}
            \includegraphics[width=0.3\textwidth]{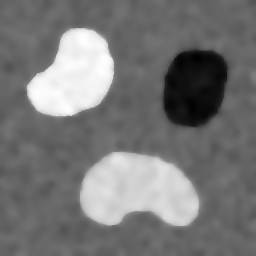} & \includegraphics[width=0.3\textwidth]{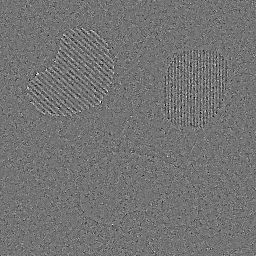} & \includegraphics[width=0.3\textwidth]{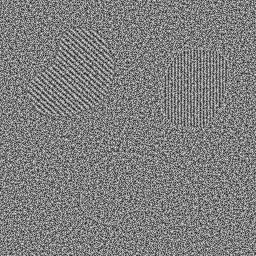} \\
            \includegraphics[width=0.3\textwidth]{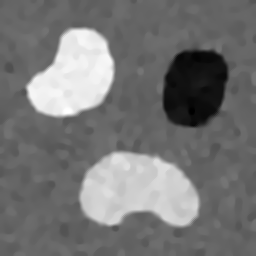} & \includegraphics[width=0.3\textwidth]{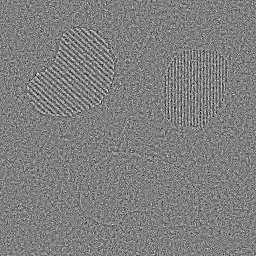} & \includegraphics[width=0.3\textwidth]{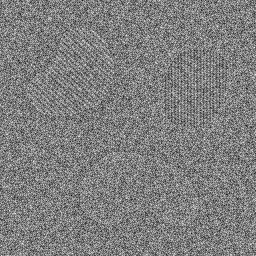} \\
            \includegraphics[width=0.3\textwidth]{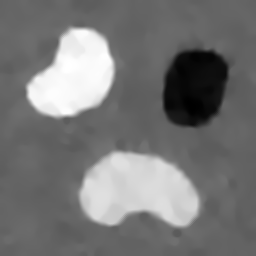} & \includegraphics[width=0.3\textwidth]{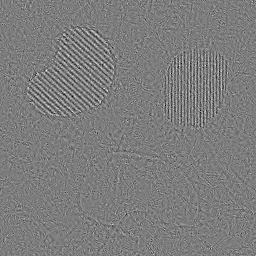} & \includegraphics[width=0.3\textwidth]{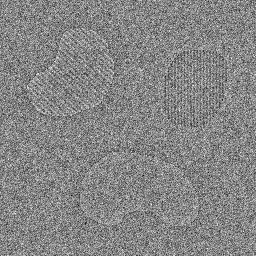} \\
        \end{tabular}
    \end{center}
    \caption{Outputs of the decomposition algorithms. First row: $F^{JG}$ algorithm; second row: $F^{AC2}$ algorithm; last row: $F^{Co}$ algorithm.}
    \label{fig:testuvwimres}
\end{figure}

Figure \ref{fig:testuvwimres} shows the outputs of the different algorithms while table \ref{tab:testuvw} gives the corresponding measures.\\

\begin{table}[!ht]
    \begin{center}
        \begin{tabular}{|m{0.2\textwidth}|m{0.2\textwidth}|m{0.2\textwidth}|m{0.2\textwidth}|} \hline
            Algorithm                         & $F^{JG}$ & $F^{AC2}$ & $F^{Co}$ \\ \hline
            $\|\tilde{u}-u_0\|_{L^2}$         & 792.8    & 873.5     & 984.6    \\ \hline
            $\|\tilde{v}-v_0\|_{L^2}$         & 1844.9   & 2832.4    & 1598.6   \\ \hline
            $\|\gamma_w-\gamma_{w_0}\|_{L^2}$ & 423.2    & 423.5     & 255.3    \\ \hline
        \end{tabular}
    \end{center}
    \caption{Evaluation Measures Obtained for All $u,v,w$ Decomposition Algorithms.}
    \label{tab:testuvw}
\end{table}

We can see the $BV$-$G$-$G$-based algorithm $F^{JG}$ has the smallest error for the structures image but the textures are slightly less preserved than the contourlet-based model $F^{Co}$. Its noisy part is of the same quality as the wavelet-based model $F^{AC2}$. Moreover, it is clear the $F^{Co}$ algorithm gives the best denoising performance and has the least residue; it also has the best score for the textures quality. Even if the visual quality seems to be close to the $F^{JG}$ algorithm, the contourlet-based model has the worst score on the structures component. Then globally, as expected, the model based on contourlet expansion gives the best decomposition.

%==============================================================================
%  CONCLUSION
%==============================================================================
\section{Conclusion}

This chapter provides an overview of structures + textures image decomposition. We also present the extension to noisy images decomposition and show that it is necessary to adopt a three-part decomposition model (structures + textures + noise). The different models are based on the bounded-variation space to describe the structures component of an image. The textures are defined by the space $G$ of oscillating functions proposed by Meyer; different stategies can be used for the noise. Some other function spaces can be chosen; most often it is equivalent to choosing the best basis or frame to represent the different components. This approach is the same philosophy as the principle of morphological component analysis recently introduced by the work of \cite{starck2,starck1}.

An interesting property used in the $BV$-$G$-$G$ model is the local adaptibility of the algorithm by choosing a nonconstant parameter $\nu$. Some recent theoretical work on the Besov and Triebel-Lizorkin spaces seems to provide some insight on the local behavior of an image (in terms of local scales). Here this approach is used to improve the quality of the decomposition.

The main problem of the decomposition models, and it remains an open question, is the choice of the different parameters. \cite{aujol2} propose a method of automatically selecting the parameter $\lambda$, but it is very expansive in computing time. We currently start some work to find some solutions.

We have proposed a method, which consists of building specific test images and using three different metrics, to evaluate the performance of the quality of components issued from the different decomposition algorithms. The first tests seem to confirm that the model based on the thresholding of contourlets coefficients is the best one. However, more complete tests based on different test images with different kind of textures, noise, or structures and by tuning the different parameters are needed. This could help us to understand completely the behaviors of this kind of algorithm.

The last topic explored in this study is the application of the image decomposition. A previous study, \cite{jegilles3}, proves that the $BV$-$G$ model enhanced the thin and long structures. Then, we use the textures component as the input of a road detection algorithm in aerial images. We believe that many applications could be created in the future.

\appendix
%==============================================================================
%  Annexes 1
%==============================================================================
\section{Chambolle's Nonlinear Projectors}\label{ap:chambolle}

\cite{chambolle} proposes an algorithm based on a nonlinear projector to solved a certain category of total variation based functional. This appendix summarizes this work. Some proofs are provided because they are relevant to the rest of the chapter.

\subsection{Notations and Definitions}
We assume the processed image is size $M \times N$.
We denote $X=\mathbb{R}^{M \times N}$ and $Y=X \times X$.

\begin{definition}
    Let $u \in X$; then the discret gradient of $u$, written $\nabla u \in Y=X\times X$, is defined by
    \begin{equation}
        (\nabla u)_{i,j}=\left((\nabla u)_{i,j}^1,(\nabla u)_{i,j}^2 \right)
    \end{equation}
    with $\forall i,j \in \llbracket 0,\ldots,M-1 \rrbracket \times \llbracket 0,\ldots,N-1 \rrbracket$
    \begin{equation}
        (\nabla u)_{i,j}^1=
        \begin{cases}
            u_{i+1,j}-u_{i,j} \qquad & \text{if} \quad i<M-1 \\[2mm]
            0 \qquad                 & \text{if} \quad i=M-1 \\
        \end{cases}
    \end{equation}

    \begin{equation}
        (\nabla u)_{i,j}^2=
        \begin{cases}
            u_{i,j+1}-u_{i,j} \qquad & \text{if} \quad j<N-1 \\[2mm]
            0 \qquad                 & \text{if} \quad j=N-1 \\
        \end{cases}.
    \end{equation}
\end{definition}

\begin{definition}
    Let $p\in Y$ $(p=(p^1,p^2))$, we define the numerical divergence operator $\Div : Y \rightarrow X$ such that $\Div=-\nabla^*$ ($\nabla^*$ is the adjoint operator of $\nabla$) by the following:
    \begin{equation}
        (\Div \; p)_{i,j}=
        \begin{cases}
            p_{i,j}^1-p_{i-1,j}^1 & \text{if} \quad 0<i<M-1 \\
            p_{i,j}^1             & \text{if} \quad i=0     \\
            -p_{i-1,j}^1          & \text{if} \quad i=M-1   \\
        \end{cases}
        +
        \begin{cases}
            p_{i,j}^2-p_{i,j-1}^2 & \text{if} \quad 0<j<N-1 \\
            p_{i,j}^2             & \text{if} \quad j=0     \\
            -p_{i,j-1}^2          & \text{if} \quad j=N-1   \\
        \end{cases}.
    \end{equation}
    We recall that $\left<-\Div \; p,u\right>_X=\left<p,\nabla u\right>_Y$.
\end{definition}

\subsection{Total Variation}
In the discrete case, the total variation can be written by:
\begin{eqnarray}
    J(u)=&\underset{0<j<N-1}{\underset{0<i<M-1}{\sum}}\left|(\nabla u)_{i,j}\right|\\
    =&\underset{0<j<N-1}{\underset{0<i<M-1}{\sum}}\sqrt{\left((\nabla u)_{i,j}^1 \right)^2+\left((\nabla u)_{i,j}^2 \right)^2}.
\end{eqnarray}
However, $J$ is a 1-homogeneous function ($J(\lambda u)=\lambda J(u)$); then if we apply the Legendre-Fenchel transform, we get:
\begin{equation}
    J^* (v)=\sup_u \left<u,v\right>_X-J(u)
\end{equation}
with
\begin{equation}
    \left<u,v\right>_X=\sum_{i,j}u_{i,j}v_{i,j},
\end{equation}
where $J^*$ is the characteristic function of the closed convex set $K$:
\begin{equation}
    J^*(v)=\chi_K(v)=
    \begin{cases}
        0       & \text{if} \quad v\in K \\
        +\infty & \text{else}
    \end{cases}.
\end{equation}
We have the property $J^{**}=J$.\\

In the continuous case (see the properties of the $BV$ space), we have:
\begin{equation}
    K=G_1=\left\{\Div \; \xi:\xi\in C_c^1(\Omega,\mathbb{R}^2);|\xi(x)|\leqslant 1,\forall x\in\Omega\right\}
\end{equation}
then
\begin{equation}
    J(u)=\sup_\xi \left\{\int_\Omega u(x)\Div \; \xi(x)dx:\xi\in C_c^1(\Omega,\mathbb{R}^2);|\xi(x)|\leqslant 1,\forall x\in\Omega\right\};
\end{equation}
however, $\int_\Omega u(x)\Div \; \xi(x)dx = \left<u,\Div \; \xi \right>_X$, then we can write:
\begin{equation}
    J(u)=\sup_\xi \left<u,\Div \; \xi \right>_X,
\end{equation}
which is equivalent, if we write $v=\Div \; \xi$, to
\begin{equation}
    J(u)=\sup_{v\in K} \left<u,v \right>_X.
\end{equation}
Now, we would like to have the same kind of expression for the discrete case. Chambolle proves the following lemma:

\begin{lem}
    In the discrete case, we have:
    \begin{equation}
        J(u)=\sup_{v\in G_1} \left<v,u\right>,
    \end{equation}
    \begin{equation}
        \text{where} \quad G_1=\left\{\Div \; p;p\in Y; |p_{i,j}|\leqslant1\right\}.
    \end{equation}
\end{lem}

\begin{definition}
    Let us define the inner product over $Y$: let $p\in Y, q \in Y$ such that $p=\left(p^1,p^2\right)$ and $q=\left(q^1,q^2\right)$; then
    \begin{equation}
        \left<p,q\right>_Y=\underset{0<j<N-1}{\underset{0<i<M-1}{\sum}}(p_{i,j}^1q_{i,j}^1+p_{i,j}^2q_{i,j}^2).
    \end{equation}
\end{definition}

\subsection{Chambolle's Projectors}
We want to solve
\begin{equation}
    \min_{u\in X} \frac{\|u-g\|^2}{2\lambda}+J(u)
    \label{eqn:fonc1}
\end{equation}

with $g\in X$, $\lambda>0$, $\|.\|$ is the euclidean norm defined by $\|u\|^2=\left<u,u\right>_X$. \\

If we apply Euler-Lagrange calculus to Eq.(\ref{eqn:fonc1}), we get

\begin{equation}
    \frac{2(u-g)}{2\lambda}+\partial J(u) \ni 0
\end{equation}
\begin{equation}
    \label{eqn:edp1}
    \Longleftrightarrow u-g+\lambda \partial J(u) \ni 0,
\end{equation}
where $\partial J$ is the subdifferential of $J$ defined by
\begin{equation}
    w\in \partial J(u) \Longleftrightarrow J(v) \geqslant J(u) + \left<w,v-u\right>_X \quad \forall v,
\end{equation}
then Eq.(\ref{eqn:edp1}) can be written as

\begin{equation}
    \frac{g-u}{\lambda} \in \partial J(u)
\end{equation}
\begin{equation}
    \Longleftrightarrow \partial J^{*}\left(\frac{g-u}{\lambda}\right) \ni u
\end{equation}
\begin{equation}
    \Longleftrightarrow \frac{u}{\lambda} \in \frac{1}{\lambda}\partial J^{*}\left(\frac{g-u}{\lambda}\right)
\end{equation}
\begin{equation}
    \Longleftrightarrow \frac{g}{\lambda} \in \frac{g-u}{\lambda}+\frac{1}{\lambda}\partial J^{*}\left(\frac{g-u}{\lambda}\right).
    \label{eqn:edp2}
\end{equation}

If we reach a minimizer of
\begin{equation}
    \frac{\left\|w-\left(\frac{g}{\lambda}\right)\right\|^2}{2}+\frac{1}{\lambda}J^{*}(w)
    \label{eqn:fonc2}
\end{equation}
then by applying Euler-Lagrange calculus to Eq.(\ref{eqn:fonc2}), we get
\begin{equation}
    w-\frac{g}{\lambda}+\frac{1}{\lambda}\partial J^{*}(w) \ni 0
\end{equation}
\begin{equation}
    \Longleftrightarrow w+\frac{1}{\lambda}\partial J^{*}(w)\ni \frac{g}{\lambda}.
\end{equation}

Thanks to Eq.(\ref{eqn:edp2}), we see that
\begin{equation}
    w=\frac{g-u}{\lambda}
\end{equation}
is a minimizer of Eq.(\ref{eqn:fonc2}).\\

However as $J^{*}(w)=\chi_{G_1}(w)$ and if $w=P_{G_1}\left(\frac{g}{\lambda}\right)$ (the projector operator over $G_1$), then $J^{*}(w)=0$ and $\left\|w-\frac{g}{\lambda}\right\|$ is minimum. We deduced that

\begin{equation}
    P_{G_1}\left(\frac{g}{\lambda}\right)=\frac{g-u}{\lambda}
\end{equation}
\begin{equation}
    u=g-\lambda P_{G_1}\left(\frac{g}{\lambda}\right).
\end{equation}
We have $P_{G_{\lambda}}\left(\frac{g}{\lambda}\right)=\lambda P_{G_1}\left(\frac{g}{\lambda}\right)$, then we have
\begin{equation}
    u=g-P_{G_{\lambda}}\left(\frac{g}{\lambda}\right).
\end{equation}

Now, we need to find how to calculate $P_{G_{\lambda}}(g)$.
Chambolle gives the following result:
\begin{equation}
    \text{computing} \; P_{G_{\lambda}}(g) \Longleftrightarrow \min_{p\in Y} \left\{\|\lambda \Div(p)-g\|^2;|p_{i,j}|^2\leqslant 1 \quad \forall i,j\right\}.
    \label{eqn:pi1}
\end{equation}
The Karush-Kuhn-Tucker conditions showed the existence of a Lagrange multiplier $\alpha_{i,j}\geqslant 0$ associated with each constraint of Eq.(\ref{eqn:pi1}) such that we have $\forall i,j$:
\begin{equation}
    -\left(\nabla\left(\lambda \Div(p)-g\right)\right)_{i,j}+\alpha_{i,j}p_{i,j}=0
\end{equation}
with
\begin{equation}
    \alpha_{i,j}>0 \quad \text{and} \quad |p_{i,j}|=1
\end{equation}
\begin{equation}
    \alpha_{i,j}=0 \quad \text{and} \quad |p_{i,j}|<1.
\end{equation}

Then we can see that if $\alpha_{i,j}=0$, then $\left(\nabla\left(\lambda \Div(p)-g\right)\right)_{i,j}=0$; which is not an interesting case. For the case $\alpha_{i,j}\neq 0$:
\begin{equation}
    \alpha_{i,j}p_{i,j}=\left(\nabla \left(\Div(p)-g\right)\right)_{i,j}
\end{equation}
\begin{equation}
    \Rightarrow |\alpha_{i,j}||p_{i,j}|=\left|\left(\nabla \left(\Div(p)-g\right)\right)_{i,j}\right|;
\end{equation}
however, $|\alpha_{i,j}|=\alpha_{i,j}$ because $\alpha_{i,j}>0$ and $|p_{i,j}|=1$; then
\begin{equation}
    \alpha_{i,j}=\left|\left(\nabla \left(\Div(p)-g\right)\right)_{i,j}\right|.
\end{equation}

Now, if we use a gradient steepest descent method with $\tau>0$; $p^0=0$; $n\geqslant 0$, we get
\begin{equation}
    p_{i,j}^{n+1}=p_{i,j}^n+\tau\left[\left(\nabla \left(\Div(p^n)-\frac{g}{\lambda}\right)\right)_{i,j}-\left|\left(\nabla \left(\Div(p^n)-\frac{g}{\lambda}\right)\right)_{i,j}\right|p_{i,j}^{n+1}\right].
\end{equation}
Finally, we get the following iterative formulation:
\begin{equation}
    p_{i,j}^{n+1}=\frac{p_{i,j}^n+\tau \left(\nabla \left(\Div(p^n)-\frac{g}{\lambda}\right)\right)_{i,j}}{1+\tau \left|\left(\nabla \left(\Div(p^n)-\frac{g}{\lambda}\right)\right)_{i,j}\right|}.
\end{equation}

Chambolle proves the following important theorem.
\begin{thm}\label{th:chambolle}
    If $\tau<\frac{1}{8}$ then $\lambda \Div(p^n)$ converges to $P_{G_{\lambda}}(g)$ when $n\rightarrow +\infty$.
\end{thm}
In pratice, we note that the choice $n=20$ is sufficient to reach the wanted convergence.

\subsection{Extension}
The previous result can be extended to the case of $BV-\mathcal{H}$ functional where $\mathcal{H}$ is a Hilbert space such that there exists a linear positive symmetric operator $K$ that defines the following norm over $\mathcal{H}$:
\begin{equation}
    \left<f,g\right>_{\mathcal{H}}=\left<f,Kg\right>_{L^2}
\end{equation}

Then, if we want to minimize
\begin{equation}\label{eq:hilb}
    J(u)+\frac{\lambda}{2}\|f-u\|_{\mathcal{H}}^2,
\end{equation}

we can use the following modified Chambolle projector:

\begin{equation}
    p_{i,j}^{n+1}=\frac{p_{i,j}^n+\tau \left(\nabla \left(K^{-1}\Div(p^n)-\frac{g}{\lambda}\right)\right)_{i,j}}{1+\tau \left|\left(\nabla \left(K^{-1}\Div(p^n)-\frac{g}{\lambda}\right)\right)_{i,j}\right|}.
\end{equation}

And the corresponding convergence theorem is shown below.
\begin{thm}\label{th:chambolle2}
    If $\tau<\frac{1}{8\|K^{-1}\|_{L^2}}$, then $\frac{1}{\lambda} K^{-1}\Div(p^n)$ converges to $\hat{v}$ when $n\rightarrow +\infty$ and $f-\frac{1}{\lambda} K^{-1}\Div(p^n)\rightarrow \hat{u}$ where $\hat{u}$ is the minimizer of Eq.(\ref{eq:hilb}).
\end{thm}

A special case is for $K=-\Delta^{-1}$, which corresponds to the Sobolev case $\mathcal{H}=H^{-1}$.

\bibliographystyle{elsarticle-num}
\bibliography{biblio2}
\end{document}